\documentclass[reprint,amsmath,amssymb,aps,floatfix]{revtex4-2}
\usepackage{xcolor}
\usepackage{graphicx}% Include figure files
\usepackage{dcolumn}% Align table columns on decimal point

\usepackage{hyperref}% add hypertext capabilities
\usepackage[mathlines]{lineno}% Enable numbering of text and display math
\usepackage{mathtools}
\usepackage{graphicx}% Include figure files
\usepackage{dcolumn}% Align table columns on decimal point
\newcommand{\be}{\begin{equation}}
\newcommand{\ee}{\end{equation}}
\newcommand{\bd}{\begin{displaymath}}
\newcommand{\ed}{\end{displaymath}}
\newcommand{\BE}{\begin{eqnarray}}
\newcommand{\EE}{\end{eqnarray}}

\newcommand{\avg}[1]{\left\langle{#1}\right\rangle}

\newcommand{\uGamma}{\underline{\Gamma}}
\newcommand{\uuLambda}{\underline{\underline{\Lambda}}}
\usepackage{comment}

\begin{document}

\title{Noisy voter models in switching environments}% Force line breaks with \\
%\thanks{green footnote to the article title}%

%%%%%%%%%%%%%%%%%%%%%%%%%%%%%%%%%%%%%%%%%%%%%%%%%%%%%

\author{Annalisa Caligiuri}
 \email{annalisa@ifisc.uib-csic.es}

\author{Tobias Galla}
    \email{tobias.galla@ifisc.uib-csic.es}
    
\affiliation{
Instituto de Física Interdisciplinar y Sistemas Complejos, IFISC (CSIC-UIB),
Campus Universitat Illes Balears, E-07122 Palma de Mallorca, Spain\\}


\date{\today}% It is always \today, today,

\begin{abstract}

%\textcolor{red}{ Lisa's comments}\\
%\textcolor{blue}{ Tobias' comments}\\
%\textcolor{teal}{ Solved}

We study the stationary states of variants of the noisy voter model, subject to fluctuating parameters or external environments. Specifically, we consider scenarios in which the herding-to-noise ratio switches randomly and on different time scales between two values. We show that this can lead to a phase in which polarised and heterogeneous states exist. Secondly, we analyse a population of noisy voters subject to groups of external influencers, and show how multi-peak stationary distributions emerge. Our work is based on a combination of individual-based simulations, analytical approximations in terms of a piecewise-deterministic Markov processes (PDMP), and on corrections to this process capturing intrinsic stochasticity in the linear-noise approximation. We also propose a numerical scheme to obtain the stationary distribution of PDMPs with three environmental states and linear velocity fields.

\end{abstract}

\maketitle

%\tableofcontents

\section{Introduction}

The voter model (VM) \cite{Clifford,Liggett75} is a model of interacting individuals, and can be used to describe, among other phenomena, the competition of opinions in a population. In the simplest setting, every agent in the population can have opinion \textit{A} or opinion \textit{B}. The individuals form an interaction network, this can be a complete graph, or the different agents can have limited sets of nearest neighbours. The interaction is an imitation process: an agent is selected at random, and adopts the opinion of a neighbour, selected randomly as well. Provided the interaction network consists of one single connected component, this model has two absorbing states, in which all agents have the same opinion (all $A$, or all $B$). These states are referred to as \textit{consensus} states. 

The voter model in this simple form was first proposed by probabilists \cite{Liggett75}, and has found widespread applications, including in the modelling of opinion dynamics, language competition and in population genetics \cite{redner2019reality,baronchelli2018emergence,kauhanen2021geospatial, HADZIBEGANOVIC20083242,ewens2004mathematical,blythe2007stochastic,FernandezGracia2014}. The VM has also generated significant interest in statistical physics, with particular focus on its coarsening dynamics \cite{ben1996coarsening,krapivsky1992kinetics}, field theoretic descriptions  and different types of phase transition and universality \cite{dornic2001critical,al-hammal2005}.

So-called `noisy' voter models (nVM) are variations of the original model. The term `noisy' is used to indicate that, in addition to the imitation process, agents can also change opinion state spontaneously. Models of this type have been used to describe people choosing among restaurants, or ants selecting one of two paths towards a source of food  \cite{Granovsky, Kirman}. The nVM has no absorbing states, and shows a finite–size phase transition \cite{Fichthorm1989,Granovsky,nVM}. When the noise is stronger than the herding mechanism the steady-state distribution is unimodal and the system displays coexistence of the two opinions. If the noise is below a threshold (set by the herding rate and the size of the population), then the stationary distribution of agents across the two opinions is bimodal. The system spends most of its time near one of the consensus states, with occasional switches from one side of phase space to the other

The models mentioned so far describe homogeneous populations in which all agents are subject to the same update rules. In \cite{galam2007role,MobiliaZealot} agents that never change opinion were introduced. These are referred to as \textit{zealots}. The effect of zealots on the VM is studied for example in \cite{MobiliaZealot,Zealots,KhalilGalla2021, vendeville2022towards}. The presence of zealots can destroy the symmetry of the steady-state distribution, and the population can become biased towards the opinion of the majority of zealots. Other, related mechanisms include the introduction of mass media \cite{mazzitello2007effects}, or personal information \cite{DeMarzo}.

The overall purpose of the present work is to study the effects of (i) time-dependence of the imitation dynamics, and (ii) time-dependent external influence on VMs. 

More specifically, with regards to (i), we study variants of the nVM in which the ratio of the noise and herding rates switches randomly between two different values. There are thus periods in which the ordering effect of herding is strong compared to disordering effect of spontaneous opinion changes, and other periods in which the disordering effects dominate. In terms of statistical physics this falls into a class of population dynamics subject to environmental fluctuations, studied for example in \cite{Assaf20131,LinDoering2016,bressloff2017stochastic,Intrinsic,Classical,Model, taitelbaum2020population, chen2023evolutionary,doering1984effect, lin2016bursting}. We also note recent work on VM in fluctuating environments \cite{mobilia2023polarization} where a three-state constrained VM under fluctuating influence is studied. Further we refer to \cite{KUDTARKAR} where the authors study a VM which switches between phases with and without noise respectively.

With regards to (ii), we introduce groups of agents who are inert to the herding mechanism (akin to zealots), but who can switch opinion states randomly from time to time. We will refer to these as \textit{influencers}. This term is to be understood broadly, in particular we do not restrict the notion of influencers to individual human actors. Instead, the term captures different types of external influences on the population of conventional VM agents, including media, advertising, social networks etc., or indeed new information, facts or events that arise and drive opinions in a population (e.g., a political scandal that comes to light). One main feature of our model is that the effects of influencers is not static, instead it fluctuates in time.  

The objective of this work is thus to understand if fluctuations of the relative noisy rate or of external influences affect the formation of consensus. At the centre of this is the question how demographic noise (due to the finiteness of the population), decision noise (random opinion changes), and external randomness interact.

To address these questions, we use a number of different approaches from statistical physics. In the limit of infinite populations, and thus discarding demographic randomness,  the system reduces to a so-called piecewise-deterministic Markov process (PDMP) \cite{PDMP,costa2008stability,azais2014piecewise,faggionato2009non}. The stationary distribution of such a process can be obtained analytically for the case of two environmental states, see for example \cite{faggionato2009non}. As a by-product of our work we develop a numerical scheme to obtain the stationary state of models with three or more environmental states. Advancing the method of \cite{Intrinsic} we also compute corrections to the infinite-population limit. This can be used to approximate the stationary distribution of the system with large but finite populations. 

Separately, analytical progress is also possible in the the adiabatic limit of fast switching environment \cite{Model}. The opposite extreme, very slow environmental changes, can also be addressed analytically.

The remainder of the paper is structured as follows. In Section~\ref{sec:Model} we define the model, including in particular the dynamics of the environment. Section~\ref{sec:Methods} contains a description of analytical approaches for very fast or very slow environmental dynamics, and, separately, in the large-population limit. In Section~\ref{sec:NVM_w_CP} 
we study a VM with fluctuating noise parameter. We obtain analytical results for fast and slow switching and we present simulation results for intermediate environmental time scales. Section~\ref{sec:NVM_w_I} focuses on the model with fluctuating influencers. 
We present our conclusions and brief outlook in Section~\ref{sec:Conclusion}.

%---------------------------------%

\section{Model definitions and methods} \label{sec:Model}
\subsection{Model definitions}
We consider a finite population of $N$ individuals. At any given time, each individual can be in one of two states, which we label as $A$ and $B$. We write $i$ for the number of individuals in state $A$, the number of individuals in state $B$ is then $N-i$. 

The composition of the population evolves in continuous time via reactions that each convert an individual of type $A$ into type $B$, or vice versa. An individual can change state through three different mechanisms: (i) they can interact with another individual and copy its state; (ii) they can change state spontaneously; or (iii) they can interact with an influencer and thus change opinion. The model operates on a fully connected graph, that is any one of the $N$ individuals can copy the state of any other individual in item (i). Similarly, the interaction with the influencers is also all-to-all, in the sense that in item (iii) any influencer can, in principle, affect any of the $N$ individuals in the population.

In order to model processes (i) and (ii) we follow the conventions of existing literature on the nVM \cite{Granovsky,nVM,Carro,KhalilGalla2021}. The external influence [process (iii)] is represented by `forces' driving the individuals towards one of the opinion states. We model these forces as a group of size $\alpha N$ (with $\alpha\geq 0$ a model parameter). We re-iterate that influencers are not necessarily to be thought of individuals, there is therefore no strict need to limit $\alpha N$ to integer values. Instead, $\alpha$ characterises the total strength of all influencers, relative that of the $N$ agents in the population. Not all influencers need to act towards the same state ($A$ or $B$). Instead, at any one time a fraction $z$ of the $\alpha N$ influencers acts towards $A$, and a fraction $1-z$ acts in the direction of opinion $B$. Naturally, $z$ is restricted to the interval $[0,1]$.

We assume that the fraction $z$ fluctuates in time. More precisely, and to allow for a compact notation, we think of the population dynamics as subject to an external environment, which can take states $\sigma=0,1,\dots,S-1$. This environment determines the fraction $z$ of influencers acting in the direction of opinion $A$ (that is $z$ is a function of $\sigma$), and it can also affect the noise rate in the dynamics. We will now describe this in detail. 

The per capita rates, in environment $\sigma$, for an agent in state $A$ to change to state $B$ and for the reverse process respectively are given by
\BE\label{eq:pi}
\pi_{A\to B,\sigma}(i)&=&a_\sigma+h\left[\frac{N-i}{(1+\alpha)N}+\frac{\alpha N(1-z_\sigma)}{(1+\alpha)N}\right], \nonumber \\
\pi_{B\to A,\sigma}(i)&=&a_\sigma+h\left[\frac{i}{(1+\alpha)N}+\frac{\alpha Nz_\sigma}{(1+\alpha)N}\right].
\EE
The quantity $a_\sigma$ is the rate of spontaneous opinion changes. 
We assume that this parameter can take different values in the different environmental states, as indicated by the subscript $\sigma$.
The coefficient $h$ is what is sometimes called a \textit{herding parameter}, and indicates how easily individuals are influenced by the opinions of other individuals, including external influencers. From the above expressions it is clear that only ratio of the noise and the herding parameters is relevant for the stationary state. We can therefore set $h=1$ throughout. This amounts to fixing the time scales of the processes in Eq.~(\ref{eq:pi}). For the time being we will keep the value of $h$ general though, as this allows us to track the origin of different terms in the dynamics.

The square brackets in the rates represent processes (i) and (iii) described above. A focal individual chooses an interaction partner either from the population of $N$ agents, or from the set of $\alpha N$ external influencers, and then adopts the opinion of this interaction partner. A change of the composition of the population occurs only if the interaction partners is in the opinion state opposite to that of the focal individual. The expression $(1+\alpha)N $ in the denominator in Eq.~(\ref{eq:pi}) is the total number of possible interaction partners, hence $(N-i)/[(1+\alpha)N]$ is the probability that the interaction partner is an individual from the population and in opinion state $B$. Similarly, $\alpha N(1-z_\sigma)/[(1+\alpha)N]$ is the probability that the interaction partner is an external influencer promoting opinion $B$.

The expressions in Eq.~(\ref{eq:pi}) are per capita rates. The total rate of converting individuals of type $B$ to type $A$ (or vice versa respectively) in the population are then
 
 \BE
 T_{i,\sigma}^+&=&(N-i)\pi_{B\to A,\sigma}(i), \nonumber \\
 T_{i,\sigma}^-&=&i\pi_{A\to B,\sigma}(i).
 \EE
 
 These are the rates with which transitions $i\to i+1$ and $i\to i-1$ occur in the population if the environment is in state $\sigma$.
 
 It remains to specify the dynamics of the environmental state. We assume that the environment undergoes a Markovian process governed by rates $\lambda\mu_{\sigma\to\sigma'}(i)$. The $\mu_{\sigma\to\sigma'}(i)$ are the elements of a stochastic matrix (with $\sum_{\sigma'} \mu_{\sigma\to\sigma'}(i)=1$ for all $\sigma$). We set $\mu_{\sigma\to\sigma}(i)=0$. In the present work the rates $\mu_{\sigma\to\sigma'}$ do not depend on the state of the population, $i$. However, to develop the general formalism we will allow for such a dependence in principle whenever possible. 
 
 The parameter $\lambda$ controls the time scale of the environmental dynamics relative to that of the changes within the population. We thus refer to the scenario $\lambda \to 0$ as the `slow-switching' limit, and to situations in which $\lambda\to\infty$ as `fast-switching'. 
 
 \subsection{Master equation}
 We write $P(i,\sigma,t)$ for the probability to find the system in state $(i,\sigma)$ at time $t$, that is the probability to have $i$ individuals of opinion $A$ in the population, and the environment in state $\sigma$. The time dependence of $P$ is omitted below to make the notation more compact. We then have the following master equation

\BE\label{eq:master}
\frac{d}{dt} P(i,\sigma)=(E-1)[T_{i,\sigma}^--1]P(i,\sigma) \nonumber
\\
+ (E^{-1}-1)[T_{i,\sigma}^+-1]P(i,\sigma) \nonumber
\\
+\lambda\sum_{\sigma'} [\mu_{\sigma'\to\sigma}(i)P(i,\sigma')-\mu_{\sigma\to\sigma'}(i) P(i,\sigma)],
\EE
where we have defined the raising operator $E$, acting on functions of $i$ as $Ef(i)=f(i+1)$. Its inverse is $E^{-1}$, i.e., we have $E^{-1}f(i)=f(i-1)$.

%-------------------------------------%
\section{Theoretical Analysis}\label{sec:Methods}

\subsection{Fast-switching limit}
In the limit of very fast environmental switching ($\lambda\to\infty$) we can, for purposes of the dynamics in the population, assume that the environmental process is at stationarity. We write $\rho_\sigma^*(i)$ for this stationary distribution. This distribution fulfills the relations

\be\label{eq:tbar}
\sum_{\sigma'}\left[ \mu_{\sigma'\to\sigma}\rho_{\sigma'}^*(i)-\mu_{\sigma\to\sigma'} \rho_\sigma^*(i)\right]=0
\ee
for all $\sigma$.

Following \cite{Classical} the dynamics of the population in the fast-switching limit is governed by effective rates
\be
\overline T_i^\pm \equiv \sum_\sigma \rho_\sigma^*(i) T^\pm_{i,\sigma}.
\ee
For our system these effective rates are

\BE\label{eq:rates_fast_switching}
\overline T_i^+&=&\left[\overline a +h\frac{i}{(1+\alpha)N}+\frac{\alpha N h\overline{z}}{(1+\alpha)N}\right](N-i) \nonumber \\
\overline T_i^-&=&\left[\overline a+ h\frac{N-i}{(1+\alpha)N}+\frac{\alpha Nh(1-\overline{z})}{(1+\alpha)N}\right] i,
\EE
where we have written 
\be\label{eq:fbar}
\overline f = \sum_\sigma \rho^*_\sigma(i)f_\sigma(i).
\ee
We have suppressed the potential $i$-dependence of objects of this type.

If model parameters are such that $\overline a\neq0$ then there are no absorbing states for this effective birth-death process. The stationary distribution is given by (see e.g. \cite{Zealots}),

\be\label{eq:stat_probability_finite_size}
\overline P_i^*=\frac{\prod_{k=1}^N \overline \gamma_{i-k}}{1+\sum_{\ell=1}^N\sum_{k=1}^\ell \overline\gamma_k},
\ee
where $\overline\gamma_i=\overline T_i^+/\overline T_{i-1}^-$. 

\subsection{Slow-switching limit}
In the slow-switching scenario, and assuming that the switching rates $\mu_{\sigma\to\sigma'}$ are not functions of $i$, the stationary distribution is given by the weighted sum of the stationary distributions $P^*(i|\sigma)$ for the system in fixed environments $\sigma\in\{0,1\}$. These distributions in turn are obtained from relations analogous to that in Eq.~\eqref{eq:stat_probability_finite_size}, but for fixed environment, and therefore with rates $T_{i,\sigma}^\pm$ instead of $\overline T_i^\pm$. We then have

\be \label{eq:SW_P}
P^*(i)=\sum_\sigma \rho_\sigma^* P^*(i|\sigma).
\ee

\subsection{Rate equations and piecewise deterministic Markov process}\label{sec:pdmp1}
\subsubsection{Piecewise deterministic Markov process in the limit of infinite populations}
In the limit of an infinite population the stochasticity within the population becomes irrelevant and a deterministic dynamics emerges between switches of the environmental state. This results in a piecewise deterministic Markov process (PDMP), see for example \cite{Intrinsic} and references therein.

Writing  $\phi=i/N$ and ${\cal T}_\sigma^\pm(i/N)= T^\pm_{i,\sigma}/N$, and taking the limit $N\to\infty$, the deterministic evolution between changes of the environment is governed by
\be
\dot \phi= {\cal T}_\sigma^+(\phi)-{\cal T}_\sigma^-(\phi).
\ee
For our model, this can be written as
\be\label{eq:pdmp}
\dot \phi = v_\sigma(\phi),
\ee
with 
\be\label{eq:flux}
v_\sigma(\phi)\equiv  a_\sigma(1-2\phi) +\frac{h \alpha}{1+\alpha}(z_\sigma-\phi).
\ee
As before, the environment $\sigma$ follows the process defined by the rates $\lambda\mu_{\sigma\to\sigma'}$.

The different terms in Eq.\eqref{eq:flux}, valid in fixed environment $\sigma$, can be interpreted as follows. The first term, $a_\sigma(1-2\phi)$, drives the population towards a state with $\phi=1/2$, i.e., equal proportions of individuals in opinions $A$ and $B$ respectively. This term describes random opinion changes, with equal rate from $A$ to $B$ or vice versa. If this was the only process in an infinite population, then a state with $\phi=1/2$ would eventually result in any fixed environment with $a_\sigma>0$. The second term on the right-hand side of Eq.\eqref{eq:flux} describes the effects of the external influencers. The fraction of influencers promoting opinion $A$ is $z_\sigma$, and a fraction $1-z_\sigma$ promotes opinion $B$. The net result of these external forces is a drive towards the state $\phi=z_\sigma$. The strength of this pull is governed by the herding parameter $h$ and by the ratio $\alpha/(1+\alpha)$ describing the strength of external influencers (of which there are $\alpha N$) among all partners a given individual can interact with ($N$ individuals in the population plus $\alpha N$ external influencers). If $h\alpha \gg (1+\alpha) a_\sigma$ then the external forces dominate the dynamics of the population, and the noise term proportional to $a_\sigma$ becomes irrelevant. 

We further note that the interaction among individuals in the population has no effect in the deterministic dynamics in Eq.\eqref{eq:flux} \cite{nVM, castellano2009statistical}. This is a well-known characteristic of the VM, and a consequence of the fact that, in an interaction of two individuals of types $A$ and $B$ respectively, the processes of individual $A$ copying opinion $B$ is equally likely as the reverse.

As a final remark, we note that the dynamics in Eq.\eqref{eq:flux} has a single attractive fixed point, given by
\be\label{eq:fixed_point}
\phi_\sigma^*=\frac{a_\sigma+h \frac{\alpha}{1+\alpha}z_\sigma}{2a_\sigma+h\frac{\alpha}{1+\alpha}}.
\ee
We always have $\phi^*_\sigma \in [0,1]$. The fixed point $\phi^*_\sigma$ is located at extreme values $0$ or $1$ only if $a_\sigma=0$, $\alpha>0$, and $z_\sigma\in\{0,1\}$. That is, for the unique fixed point to be at $0$ or $1$, there must not be any spontaneous opinion changes, there must be a non-zero set of influencers, and all influencers must act in the same direction. 

Further, most of our paper excludes cases in which two different environmental states lead to the same fixed point, i.e., we assume that $\phi^*_\sigma\neq \phi^*_{\sigma'}$ for $\sigma\neq\sigma'$ and $\alpha, h$ $\neq 0$. 
Without loss of generality we can then assume that the environmental states $\sigma=0,\dots,S-1 $ are labelled such that $\phi_0^*<\phi_1^*<\dots<\phi_{S-1}^*$. The dynamics of the PDMP is then restricted to the interval $(\phi_0^*,\phi_{S-1}^*)$, where $\phi_0^*$ is the left-most fixed point, and $\phi^*_{S-1}$ is the right-most fixed point on the $\phi$-axis.

\subsubsection{Stationary distribution}
The PDMP defined in Sec.~\ref{sec:pdmp1}, governed by Eqs.~(\ref{eq:pdmp}) and the dynamics of the environmental process can be described by the following Liouville-master equation for the probability $\Pi(\phi,\sigma)$ to find the system in state $(\phi,\sigma)$,

\BE\label{eq:liouville}
\frac{d}{dt} \Pi(\phi,\sigma)=-\frac{\partial}{\partial \phi} \left[ v_\sigma(\phi)\Pi(\phi,\sigma)\right] \nonumber
\\
+\lambda\sum_{\sigma'} [\mu_{\sigma'\to\sigma}(\phi) \Pi(\phi,\sigma')-\mu_{\sigma\to\sigma'}(\phi) \Pi(\phi,\sigma)].
\EE
In slight abuse of notation we have written $\mu_{\sigma\to\sigma'}(\phi)$ for the transition rates of the environmental process if the population is in state $\phi$.

The stationary state of the PDMP is defined by $\frac{d}{dt}\Pi(\phi,\sigma)=0$ for all $\phi,\sigma$. In this state we have
\BE \label{eq:pdmp_equation}
\frac{\partial}{\partial \phi} \left[ v_\sigma(\phi)\Pi(\phi,\sigma)\right]
&=&\lambda\sum_{\sigma'} \left[\mu_{\sigma'\to\sigma}(\phi) \Pi(\phi,\sigma')\right.\nonumber \\
&&\left.-\mu_{\sigma\to\sigma'}(\phi) \Pi(\phi,\sigma)\right].\label{eq:pdmp_stat}
\EE

%-------------%

\subsubsection{Special case of two environmental states}\label{sec:pdmp_2states}
If there are only two environmental states, $\sigma\in\{0,1\}$ then the stationary state can be found explicitly \cite{Intrinsic, faggionato2009non}, and takes the following form,
\BE\label{eq:2states_stat}
\Pi(\phi,0)&=&\frac{{\cal N}}{-v_0(\phi)}g(\phi), \nonumber \\
\Pi(\phi,1)&=&\frac{\cal N}{v_1(\phi)}g(\phi),
\EE
where $\phi\in(\phi_0^*,\phi_1^*)$, and
\be
g(\phi)=\exp\left[-\lambda \int^\phi du \left(\frac{\mu_{0\to 1}(u)}{v_0(u)}+\frac{\mu_{1\to 0}(u)}{v_1(u)}\right)\right].
\ee
We note that $v_0(\phi)<0$ and $v_1(\phi)>0$ for $\phi\in(\phi_0^*,\phi_1^*)$. The constant ${\cal N}$ in Eq.~(\ref{eq:2states_stat}) is determined by normalisation, $\int_{\phi_0^*}^{\phi_1^*} du \left [\Pi(u,0)+\Pi(u,1)\right]=1$.

\subsubsection{Systems with more than two environmental states}

For systems with three or more environmental states we do not know of any method to find the stationary distribution of the resulting PDMP analytically. However, is possible to numerically integrate the system in Eq.\eqref{eq:pdmp_equation}.

To deal with singularities in Eq.~\eqref{eq:pdmp_equation} at the fixed points $\phi_\sigma^*$ one can  divide the interval $0<\phi<1$ into $S-1$ subintervals $\phi_\sigma^*<\phi<\phi_{\sigma+1}^*$ ($\sigma=0,\dots,S-2$), and perform a numerical integration in each of these intervals. One then needs to ensure  continuity of all functions $\Gamma_\sigma(\phi) = v_\sigma(\phi)\Pi(\phi,\sigma)$ at the boundaries. Further details can be found in Appendix \ref{App:Algorithm for three state of the influencers}.

\subsection{Leading-order effects of noise}
The PDMP descriptions retains the environmental noise, but discards all intrinsic stochasticity at fixed environmental state. This approach is formally valid in the limit of infinite populations, $N\to\infty$. The effects of noise within the population can be studied to leading order by an expansion of the master equation (\ref{eq:master}) in powers of $1/N$. This follows the lines of \cite{Intrinsic}.

To leading order the expansion produces the PDMP, and to sub-leading order a dynamics described by a set of `piecewise' stochastic differential equations is obtained \cite{Intrinsic,Classical, hufton2019model}. More precisely, these are of the form $\dot x = v_\sigma(x)+\sqrt{w_\sigma(x)/N}\eta(t)$, where $\eta$ is zero-average Gaussian white noise of unit amplitude (i.e., $\avg{\eta(t)\eta(t')}=\delta(t-t')$). The functions $w_\sigma(x)$ are given by \cite{Intrinsic}
\be
w_\sigma(x)={\cal T}^+_\sigma(x)+{\cal T}_\sigma^-(x).
\ee
   As before, the environmental state undergoes the Markov process defined by the rates $\lambda\mu_{\sigma\to\sigma'}$.

As shown in \cite{Intrinsic}, further progress can then be made using a linear-noise approximation. To this end, one writes $i/N=\phi+N^{-1/2}\xi$, where $\phi$ is the trajectory of the PDMP for a given realisation of the environmental dynamics [i.e., $\dot\phi=v_\sigma(\phi)$]. Expanding to linear order in $\xi$ one then finds
\be
\dot \xi(t)=v_\sigma'(\phi)\xi+\sqrt{w_\sigma(\phi)}\zeta(t)
\ee
with white Gaussian noise $\zeta(t)$, and where $v_\sigma'(\phi)=dv_\sigma(\phi)/d\phi$. 
The stationary distribution of the original system in Eq.~(\ref{eq:master}) can be approximated by the following expression,
\BE\label{eq:stat_distribution_with_noise}
\Pi(x)&=&\sum_\sigma\int\, d\phi\, d\xi\, \left[\Pi(\phi,\sigma)\Pi(\xi|\phi)\right. \nonumber \\
&&\times\left. \delta(x-\phi-N^{-1/2}\xi)\right].
\EE
Here, $\Pi(\phi,\sigma)$ is the stationary distribution of the PDMP, and $\Pi(\xi | \phi)=[2\pi s^2(\phi)]^{-1/2}\exp\left[-\xi^2/[2s^2(\phi)]\right]$ is a Gaussian distribution with mean zero and variance given by
\be\label{eq:s2_1}
s^2(\phi)=-\frac{1}{2}\frac{\sum_\sigma \Pi(\sigma|\phi)w_\sigma(\phi)}{\sum_\sigma \Pi(\sigma|\phi)v_\sigma'(\phi)}.
\ee
This relation was derived for systems with two environmental states in \cite{Intrinsic}, but holds more generally as described in more detail in Appendix \ref{App:Variance}. 

%%%%%%%%%%%%%%%%%%%%%%%%%%%%%%%%%%%%%%%%

\section{Noisy voter model with switching noise parameter}\label{sec:NVM_w_CP}
\subsection{Setup}
In this section, we will examine the simple case of $\alpha=0$, i.e., the system is not affected by any influencers. The rates in environmental state $\sigma$ are then
\BE\label{eq:rates_switch_h}
T_{i,\sigma}^+&=&\left(a_\sigma+h\frac{i}{N}\right) (N-i),\nonumber \\
T_{i,\sigma}^-&=&\left(a_\sigma+h\frac{N-i}{N}\right) i.
\EE

Despite the absence of influencers the system operates within a switching environment, as the noise parameter $a_\sigma$ fluctuates in time.
We study the case of two environmental states $\sigma=0,1$. We label the states such that $a_0<a_1$.
The rates for the environmental switches in our analysis are assumed not to depend on the population state $i$. Therefore, the stationary distribution for the environmental state $\sigma$ will be simply
\be\label{eq:stat_env_2states}
\rho_0^*=\frac{\mu_{1\to 0}}{\mu_{1\to 0}+\mu_{0\to 1}}, ~~ \rho_1^*=\frac{\mu_{0\to 1}}{\mu_{1\to 0}+\mu_{0\to 1}}.
\ee

Throughout our analysis we assume $\mu_{0\to 1}=\mu_{1\to 0}$, and consequently we have $\rho^*_0=\rho^*_1=1/2$.

We will first investigate the slow-switching and fast-switching limits. The total rate for events in the population in environment $\sigma$ is $T_{i,\sigma}^++T_{i,\sigma}^-=a_\sigma N + 2h i (N-i)/N$, and therefore takes values between $a_\sigma N$ and $(a_\sigma+h/2)N$. The environment can therefore be considered slow when $\lambda\ll a_0N$.  Similarly, the environment is fast relative the population when $\lambda \gg N(a_1+h/2)$. 

\subsection{Slow switching limit}\label{sub:SlowSwitchCP}

\begin{figure}
    \centering        
    \includegraphics[width=0.5\textwidth]
    {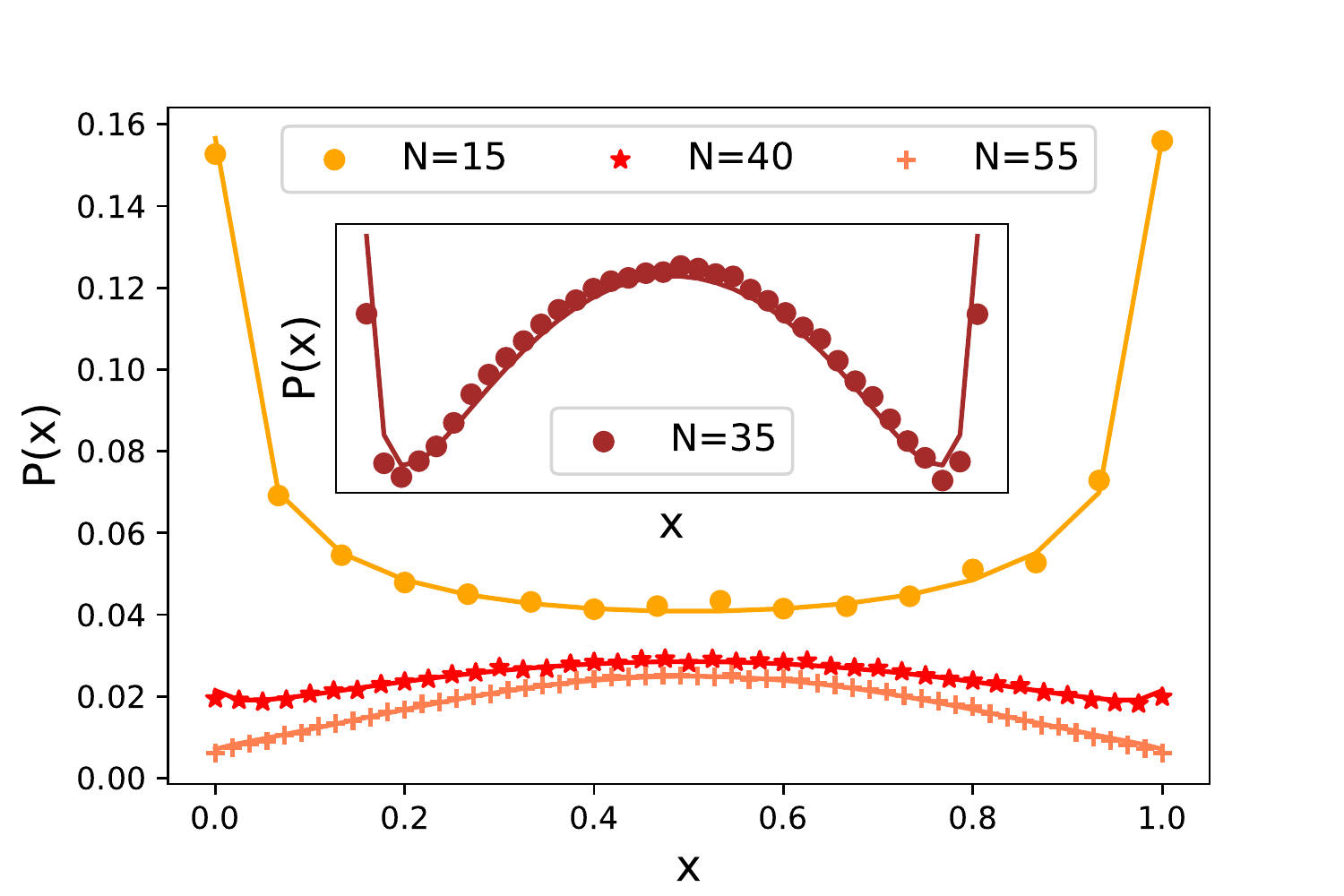}   
    
    \caption{{\bf Voter model with slow-switching noise rate.} Stationary distribution from simulations (symbols) and from theory 
    [lines, from Eq.~\eqref{eq:SW_P}]. Model parameters are $a_0=0.02$, $a_1=0.05$, $h=1$ and $\lambda=0.02$. In the inset, we highlight the new trimodal shape ($N=35$). Each distribution is from $10^6$ entries sampled every $50$ units of time, after an initial transient of $1000$ units of time}
     \label{fig:CP_Slow}
\end{figure}

We show the stationary distribution in the slow-switching limit in Fig.~\ref{fig:CP_Slow}, comparing theoretical predictions with simulation results for different values of the population size $N$. 

We observe three different shapes. For small populations ($N=15$ in the figure) the distribution is bimodal, with two maxima at the consensus states. When the population is large ($N=55$) we find a unimodal shape, the population is mostly in states in which both opinions coexist in similar proportions. 

\begin{figure}
    \includegraphics[width=0.5\textwidth]{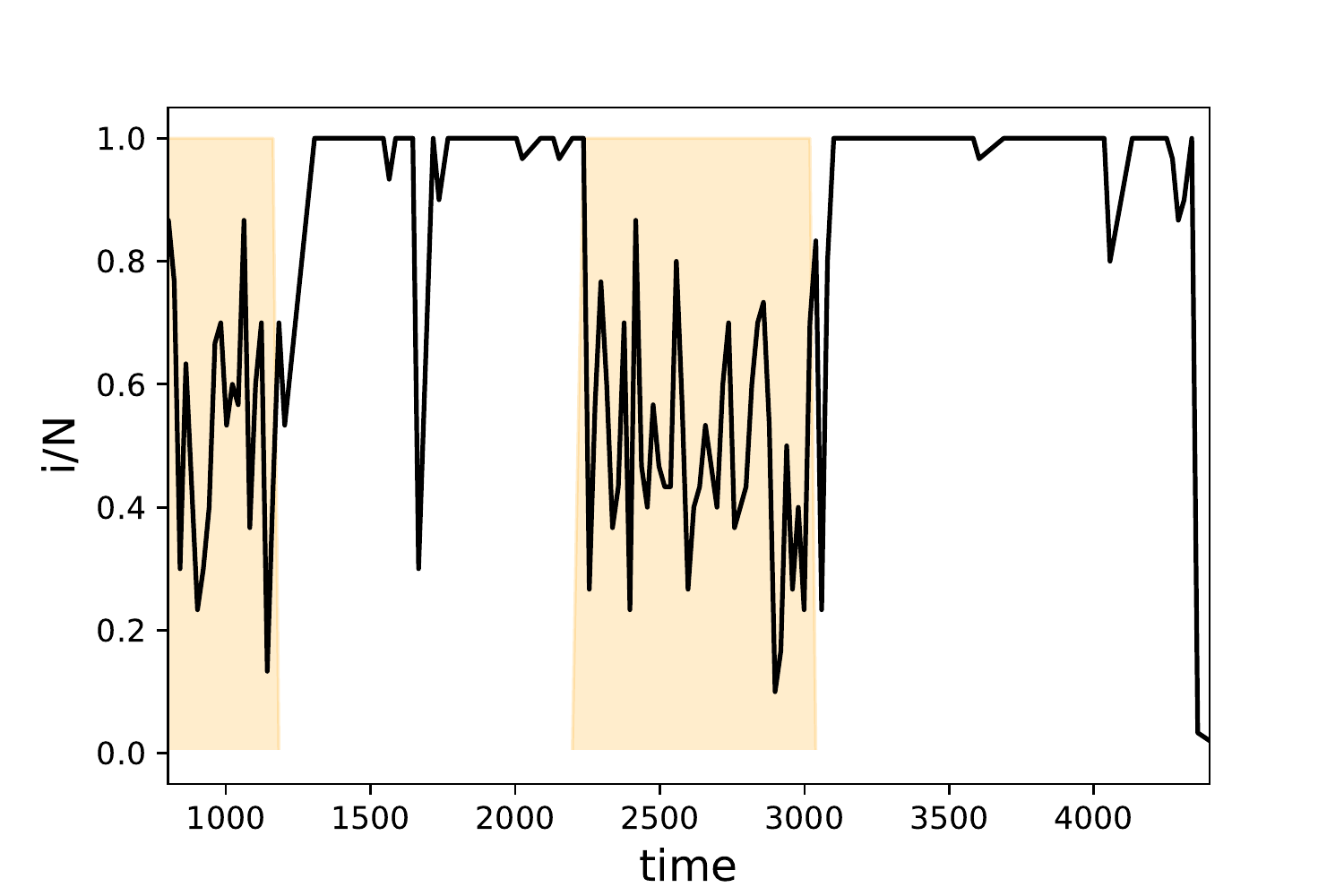}
    \caption{{\bf Time series of the fraction of agents in state $A$ from a simulation of the voter model with switching noise parameter}. Shaded segments indicate high noise rate, white background low noise rate. Model parameters are $a_0=0.001$, $a_1=0.1$, $h=1$, $\lambda=0.001$ and $N=30$.}
     \label{fig:time_series}
\end{figure}

So far, this is similar to what one would expect in the conventional two-state VM, namely transition from a bimodal shape in small populations to a unimodal shape in large populations \cite{nVM}. However, in the present model we find an additional phase with trimodal distributions for intermediate population sizes ($N=40$ in Fig.~\ref{fig:CP_Slow}). The distribution has two maxima at the extremes, and an additional maximum in the center. This state is characterized by alternating periods of coexistence of both opinions and periods of polarization. This is illustrated by the time series in Fig.~\ref{fig:time_series}. Broadly speaking this type of behaviour represents a scenario in which public opinion is characterized by a mixture of two views, but where event may occur temporarily increasing the weight of herding relative to that of noise, and thus polarising opinions.

\subsection{Fast switching limit}\label{sub:FastSwitchCP}

In the limit of fast environmental switching we have effective transition rates
\BE\label{eq:rates_switch_h_fs}
\overline T_{i}^+&=&\left(\overline a+h\frac{i}{N}\right) (N-i),\nonumber \\
\overline T_{i}^-&=&\left(\overline a+h\frac{N-i}{N}\right) i
\EE

\begin{figure}
    \centering
    \includegraphics[width=0.5
    \textwidth]{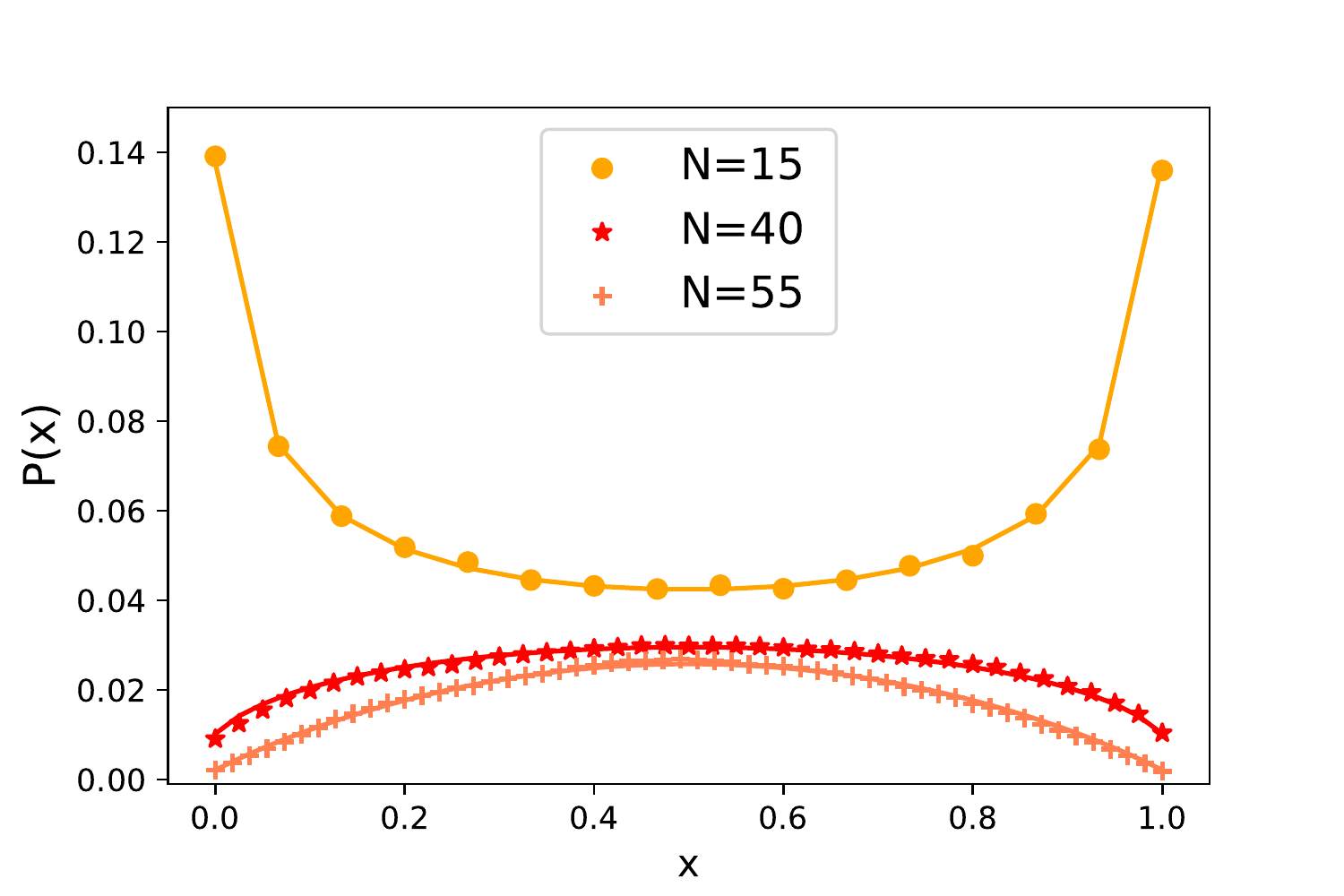}
    \caption{{\bf Voter model with fast-switching noise rate.} Stationary distribution from simulations (symbols), and from theory [lines, Eq.~\eqref{eq:stat_probability_finite_size}, with noise parameter $\overline a=(a_0+a_1)/2$]. Model parameters: $a_0=0.02$, $a_1=0.05$, $h=1$ and $\lambda=100$. Each distribution is from $5\times 10^6$ samples; time between subsequent samples is $\Delta t=0.01$, after a transient of $500$ units of time.}
    \label{fig:CP_Fast}
\end{figure}

This describes a conventional noisy VM \cite{nVM}, with noise parameter $\overline a$ and herding parameter $h$. The stationary distribution is bimodal if $N< h / \overline a$ , and unimodal otherwise, as shown in Fig.~\ref{fig:CP_Fast}. The transition between these two regimes occurs without an intermediate trimodal shape.

\subsection{Simulations for intermediate switching rates}

When the time scales for population and environmental switches are comparable to each other, an analytical characterisation is not easily available. Nonetheless, we can conduct simulations, varying the value of $\lambda$ to interpolate between the slow-switching regime in Sec.~\ref{sub:SlowSwitchCP} to fast switching in Sec.~\ref{sub:FastSwitchCP}.

\begin{figure}
    \includegraphics[width=0.5\textwidth]{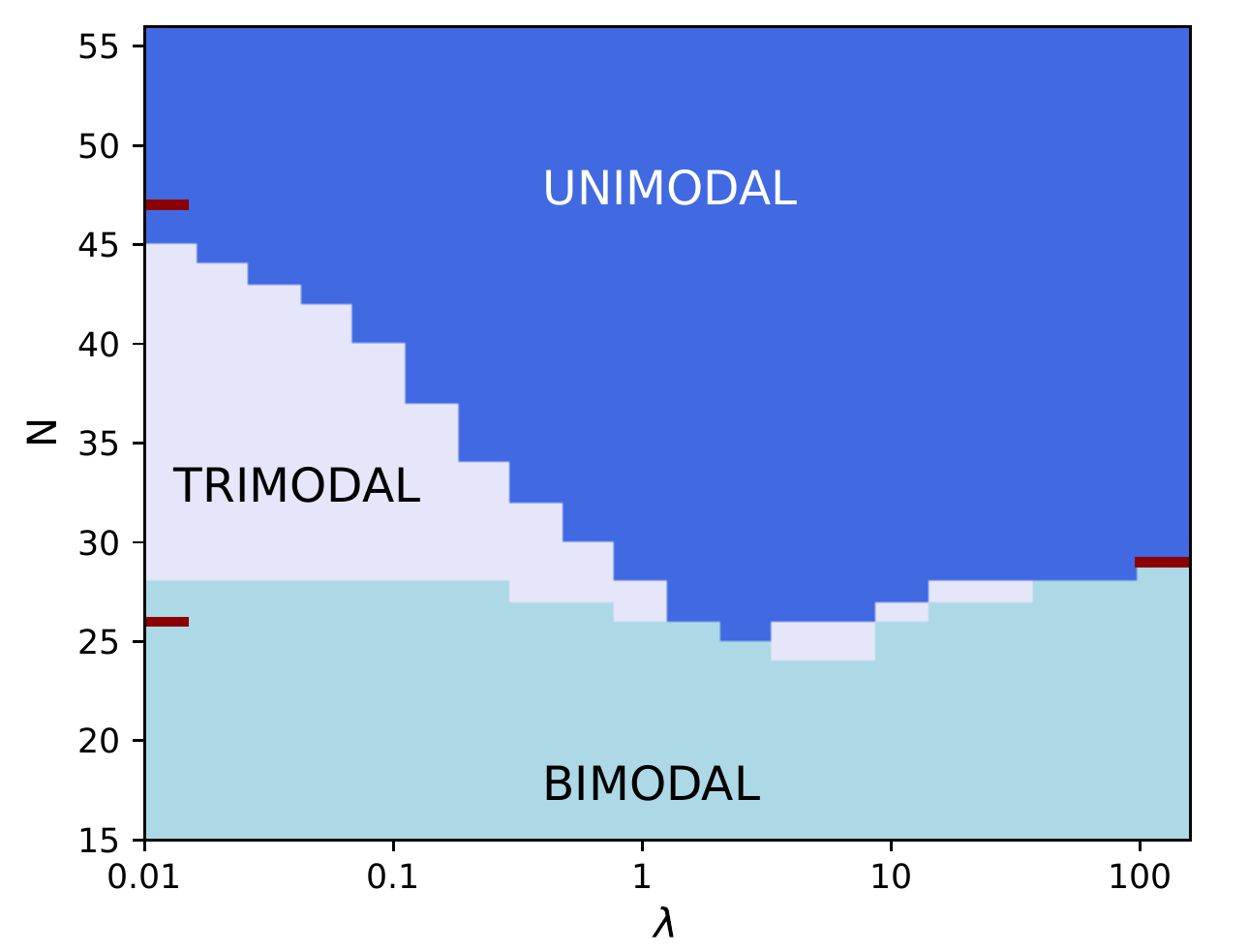}
    \caption{{\bf Phase diagram for the voter model with switching noise parameter}. The coloured shading indicates the shape of the stationary distribution as found in simulations. The red lines on the left and right show the phase boundaries in the limits of slow switching [left, found from  evaluating the expression in Eq.~\eqref{eq:SW_P}], and fast switching [right, from Eq.~\eqref{eq:stat_probability_finite_size}]. For each pair of values for $N$ and $\lambda$, we obtain the stationary distribution $P^*(i)$ ($i=0,\dots,N$). Using the expected symmetry $P^*(i)=P^*(N-i)$, we classify a distribution as unimodal when $P^*(0)<P^*(1)$, as trimodal when $P^*(0)>P^*(1)$ and when there is a local maximum in the interval $\left[N/2-1,N/2+1 \right]$, and as bimodal otherwise.
    Model parameters are $a_0=0.02$, $a_1=0.05$, $h=1$.
    Time between subsequent samples is $\Delta t=1/\lambda$, for each distribution we take $10^5-10^7$ samples after a transient of $500$ units of time. }
     \label{fig:CP_phase_diagram}
\end{figure}

Figure \ref{fig:CP_phase_diagram} shows the resulting phase diagram in the ($\lambda, N)$-plane, at fixed values of the remaining model parameters. For slow switching (low values of $\lambda$), the stationary distribution exhibits three different shapes (bimodal, trimodal, and unimodal) as $N$ increases. For faster environmental dynamics (higher values of $\lambda$), the trimodal shape disappears, resulting in the well-known finite-size transition between unimodal and bimodal states in a nVM with an effective noise constant.

%%%%%%%%%%%%%%%%%%%%%%%%%%%

\section{Noisy voter model with switching influencers}\label{sec:NVM_w_I}

In this section we focus on the impact of fluctuating groups of influencers on the nVM. The state of the influencers plays the role of the external environment. We assume that $a_\sigma\equiv a$ and $h_\sigma \equiv 1$ across environmental states. We begin by examining the two-state scenario, which allows us to obtain an explicit solution for stationary distribution of the model. If there are more than two environmental states we resort to numerical integration to solve Eq.~\eqref{eq:pdmp_stat}.

\subsection{Two states of the group of influencers}
We consider a model with two environmental states and in which all influencers form one group of total strength $\alpha N$. At any one time, they act coherently either in favour of opinion $A$ or of opion $B$. As before we write $\sigma\in\{0,1\}$ for the two environmental states, and $\lambda\mu_{0\to 1}$ and $\lambda\mu_{1\to 0}$ for the switching rates. We have $z_0=0$ and $z_1=1$. The stationary state of the environmental dynamics is  given by Eq.~(\ref{eq:stat_env_2states}).

\subsubsection{Fast-switching limit}
We first consider the fast-switching limit $\lambda\to\infty$. 
The effective rates in Eq.~\eqref{eq:rates_fast_switching} then become
\BE\label{eq:eff_rates}
\overline T_i^+ &=&\left[a^++\frac{i}{(1+\alpha)N}\right](N-i), \nonumber \\
\overline T_i^- &=&\left[a^-+\frac{N-i}{(1+\alpha)N}\right]i,
\EE
where we have introduced
\BE
a^+&\equiv& a+\frac{\alpha  \overline z}{1+\alpha}, \\
a^-&\equiv& a+\frac{\alpha  (1-\overline z)}{1+\alpha},  
\EE
with $\overline z=\rho_0^* z_0+\rho_1^*z_1$ [see also Eq.~(\ref{eq:fbar})].

We note that Eqs.~\eqref{eq:eff_rates} are also valid for an arbitrary number of environmental states (with the definition $\overline z = \sum_\sigma \rho_\sigma^*(i) z_\sigma$, and so long as $h=1$ and $a_\sigma\equiv a$).
Eqs.~\eqref{eq:eff_rates} are recognised as the transition rates of a potentially asymmetric noisy voter model (asymmetry here refers to setups with $a^+\neq a^-$). 

\begin{figure}
    \centering
    \includegraphics[width=0.5\textwidth]{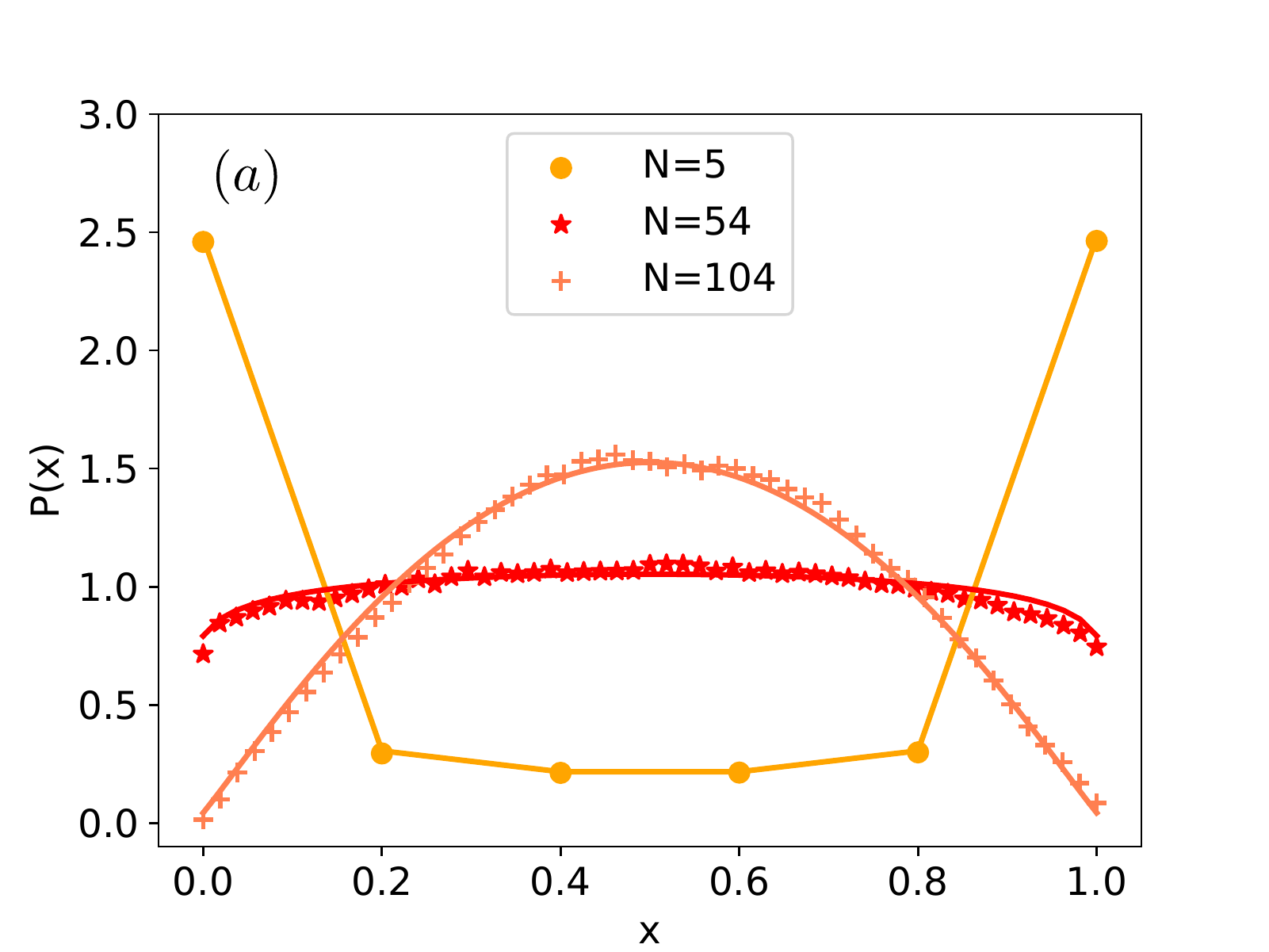}
    \includegraphics[width=0.5\textwidth]{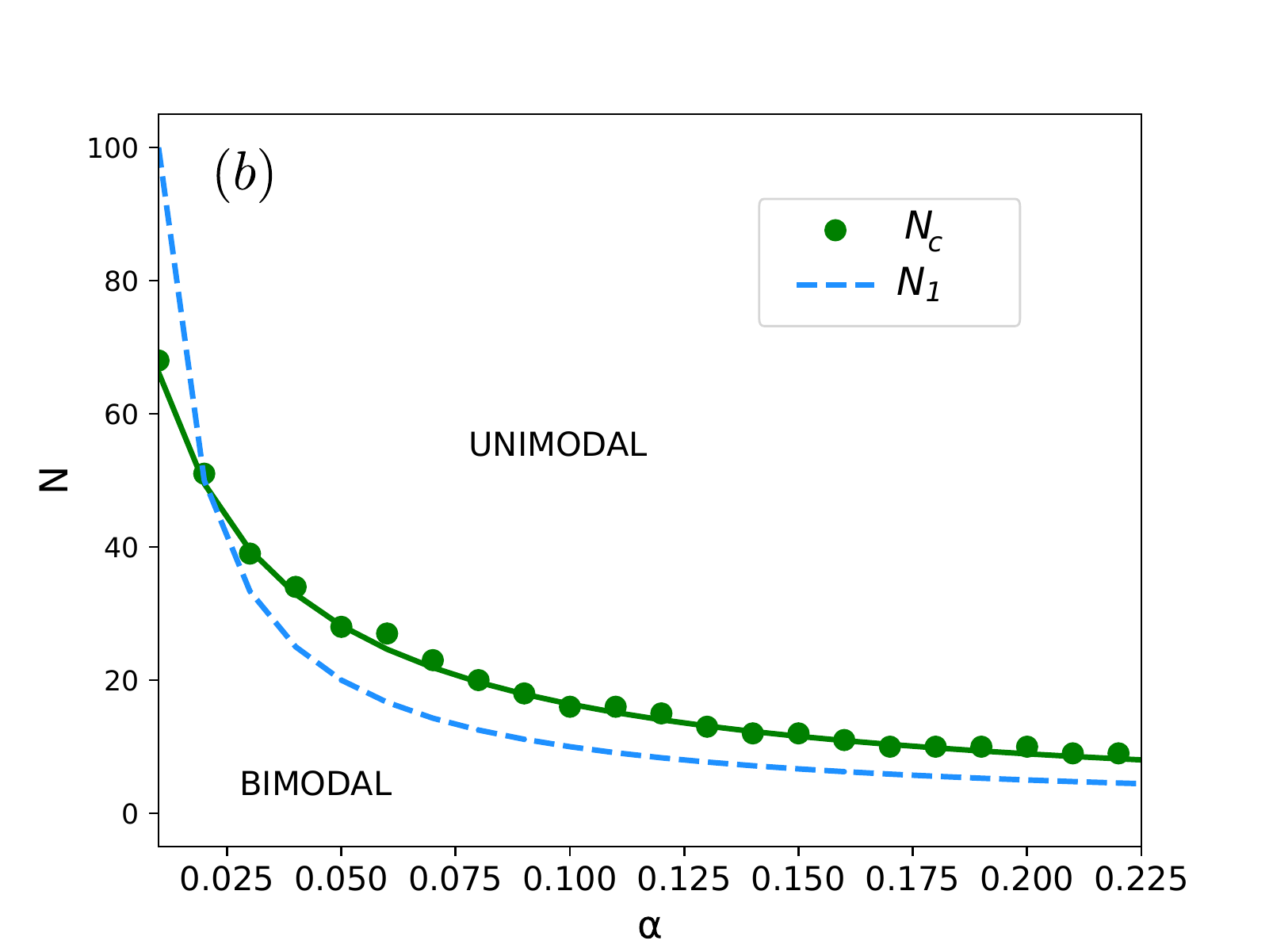}
    \caption{{\bf Transition between bimodal and unimodal stationary distributions in the model with two external states and fast switching influencers.} 
    Panel (a): Stationary distributions for different values of $N$ and $\alpha=0.02$.
    For $N=5$ the stationary distribution is bimodal; for $N=54$ and $N=104$ it is unimodal. Panel (b): Location of the phase transition, $N_c(\alpha)$, as a function of the weight $\alpha$ of the influencers. The prediction from Eq.~\eqref{eq:n_c} is shown as a solid line,  markers are from simulations.
    Below $N_c(\alpha)$ the stationary distribution has a bimodal shape, above it is unimodal. The dashed line shows $N_1=1/\alpha$. Below this line the total weight of influencers is less than that of one normal agent. Model parameters are $\lambda\mu_{0\to 1}=\lambda_{\mu_1\to 0}=50$, $z_1=1-z_0=0$, $a=0.01$.
    Time between subsequent samples is $\Delta t=0.1$, we take $10^6$ samples after a transient of one unit of time. 
}
    \label{fig:f_s_2_state_sym}
\end{figure}

For $\overline z=1/2$ one has a symmetric noisy voter model with effective herding rate $1/(1+\alpha)$ and with noise parameter $a^+=a^-=a+\alpha/[2(1+\alpha)]$. A finite-size transition between unimodal and bimodal states occurs in the nVM when the ratio of noise parameter to herding parameter is $1/N$ \cite{nVM,castellano2009statistical,Consensus}. This leads to
\be\label{eq:n_c}
N_c =\frac{1}{a(1+\alpha)+\alpha/2}.
\ee
Simulations results verifying this are shown in Fig.~\ref{fig:f_s_2_state_sym}.

The total weight of influencers in the model is the equivalent of $\alpha N$ normal agents. For a given value of $\alpha$ this means that the weight of influencers is less than that of a single normal agent when there are fewer than $N_1\equiv 1/\alpha$ normal agents ($N<1/\alpha$). In such situations one cannot think of influencers as discrete agents. 
 
We now briefly consider the asymmetric case, $\overline z \neq 1/2$. In this case, the stationary distribution is no longer symmetric [i.e., the distribution will not fulfill $P^*(i)=P^*(N-i)$ for all $i$]. We therefore study the shape of the distribution near its left and right ends of the domain separately. As parameters are varied, the `slope' of the distribution near the left end changes when $P(i=0)=P(i=1)$. This is the case if and only if $\overline T_0^+=\overline T_1^-$. This in turn leads to 

\be\label{eq:n_left}
a(1+\alpha)(N-1)+\alpha\left[N-\overline z(N+1)\right]-\frac{N-1}{N}=0.
\ee

For given $a, \alpha$ and $\overline z$ we denote the physically relevant solution of this equation by $N_c^{\rm left}$. An analogous equation is obtained from setting $\overline T_N^-=\overline T_{N-1}^+$,

\be\label{eq:n_right}
a(1+\alpha)(N-1)+\alpha\left[\overline z(N+1)-1\right]-\frac{N-1}{N}=0.
\ee

\begin{figure}
    \centering
    \includegraphics[width=0.5\textwidth]{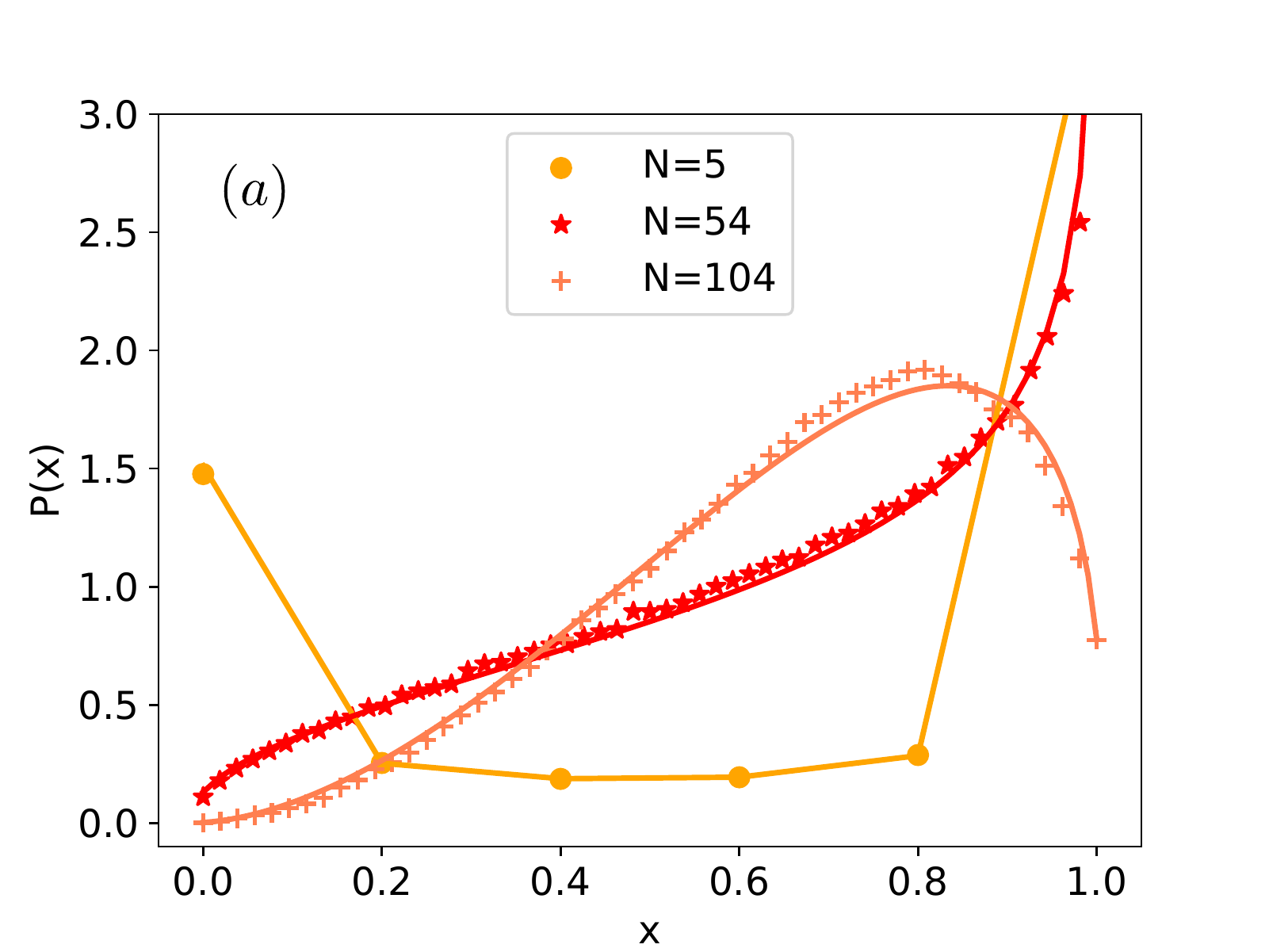}
    \includegraphics[width=0.5\textwidth]{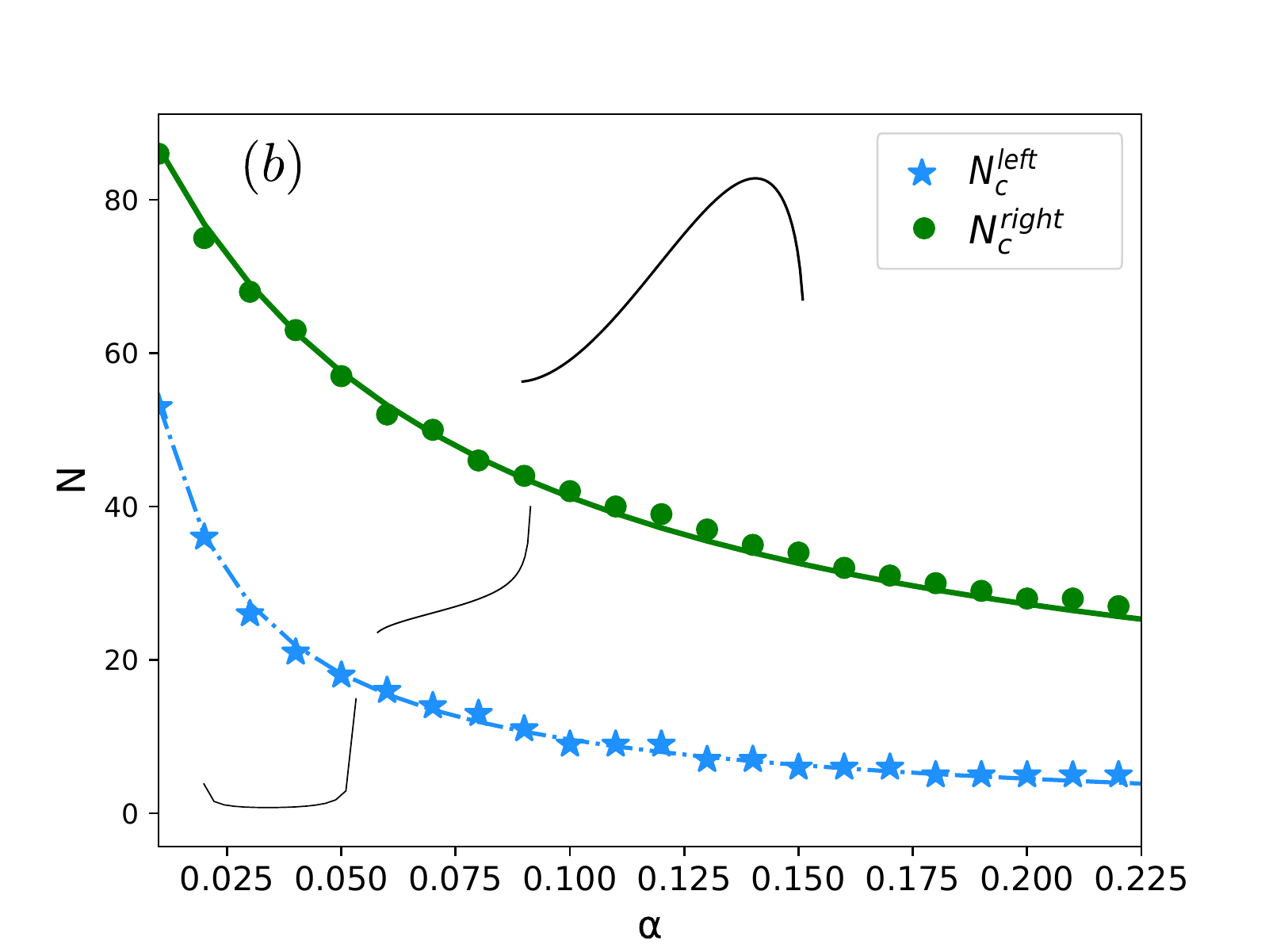}
    
    \caption{{\bf Model with asymmetric influencers in the fast-switching limit.} Panel (a) shows  the shapes of the stationary distribution for $x$ in a model with two environmental states, $\overline z =0.85$ and fast switching, for different sizes of the population. Remaining model parameters are $a=0.01$, $\alpha=0.02$. 
    Markers are from simulations, lines from the analytical theory in the fast-switching limit. Time between subsequent samples in simulations is $\Delta t = 0.1$, we take $10^6$ samples after a transient of one unit of time. 
   Panel (b) shows $N_c^{\rm left}$ and $N_c^{\rm right}$ from Eqs.~\eqref{eq:n_left} and \eqref{eq:n_right} respectively (lines). Markers are from simulations. 
    For $N<N_c^{\rm left}$ the distribution is bimodal and asymmetric, in the area between the lines it is strictly increasing in $x$, and for $N>N_c^{\rm right}$ the distribution has a unimodal asymmetric shape.}\label{fig:s_2_state_asym}
\end{figure}

We denote the solution of this equation by $N_c^{\rm right}$.
The resulting behaviour of the asymmetric model is illustrated in Fig.~\ref{fig:s_2_state_asym} (a). In the example shown we have $\overline z=0.85$ so that influencers tend to favour opinion $A$. In this setup one finds $N_c^{\rm left}<N_c^{\rm right}$. For relatively small populations ($N<N_c^{\rm left}$) the stationary distribution is bimodal, but with a higher peak at $x=1$ than at $x=0$. As $N$ is increased, the left edge of the distribution (near $x=0$) first changes slope, and a distribution which is strictly increasing in $x$ results for $N_c^{\rm left}<N<N_c^{\rm right}$. Finally, when $N>N_c^{\rm right}$ the distribution is unimodal, but with its maximum closer to $x=1$ than to $x=0$. 
Fig.~\ref{fig:s_2_state_asym}(b) shows the resulting phase diagram in the $(\alpha,N)$ plane, indicating the transitions between a bimodal phase in small populations, a phase with a strictly increasing functional shape for the stationary distribution at intermediate population sizes, and finally a unimodal phase for large populations.

\subsubsection{Limit of large populations}
In the limit $N\to\infty$ the internal noise in the population becomes irrelevant, and a PDMP results. The velocities in the two environments are given in Eq.~(\ref{eq:flux}). Using the expressions in Sec.~\ref{sec:pdmp_2states} the stationary distribution for the model with two environmental states can be obtained for any choice of the rates $\lambda \mu_{0\to 1}$ and $\lambda\mu_{1\to 0}$. We here restrict the discussion to the case $\mu_{0\to 1}=\mu_{1\to 0}=1$, but keep the time scale separation $\lambda$ general. We then find

\be\label{eq:2S_pdmp_stat}
\Pi^*(\phi)={\cal C}\left[(\phi-\phi_0^*)(\phi_1^*-\phi)\right]^{\lambda/\lambda_c-1},
\ee
where ${\cal C}$ is a normalisation constant, and where
\be\label{eq:lambda_c_2s_pdmp}
\lambda_c=2a+\frac{\alpha}{1+\alpha}.
\ee
The fixed points $\phi_0^*$ and $\phi_1^*$ are obtained from Eq.~\eqref{eq:fixed_point}. The stationary distribution becomes singular at $\phi=\phi_0^*$ and $\phi=\phi_1^*$ respectively for $\lambda<\lambda_c$.

An example is shown in Fig.~\ref{fig:stat_distrib_2_state_N200}. For $\lambda<\lambda_c$ the distribution is bimodal as shown in panel (a). For $\lambda=\lambda_c$ the distribution is mostly flat [panel (b)], and for $\lambda>\lambda_c$ a unimodal state results [panel (c)].

These results can be understood from the form of the flow fields $v_\sigma(\phi)=a(1-2\phi)+\alpha (z_\sigma-\phi)/(1+\alpha)$ obtained from Eq.~\eqref{eq:flux}. In each environment $\sigma$, the variable $\phi$ thus moves towards the fixed point $\phi_\sigma^*$ on a characteristics time scale given by $[2a+\alpha/(1+\alpha)]^{-1}$. The inverse of this time scale sets the value $\lambda_c$ for the switching rate, separating the unimodal and bimodal regimes. Thus, for $\lambda<\lambda_c$ the environmental switching is slower than the relaxation of the population in any fixed environment. This relaxation can therefore proceed before the next switch occurs, and hence probability accumulates near the fixed points. The distribution of $\phi$ is bimodal, and if inspected at a given time, the population is likely to be found near the consensus state favoured by the influencers in the environmental state at that time. 

If on the other hand $\lambda>\lambda_c$ then the environment switches quickly, before the population can approach either fixed point. The system frequently reverses its direction of motion, and the most likely states of the variable $\phi$ are those in the interiour of the interval from $0$ to $1$. As a result, the stationary distribution is peaked in the middle (unimodal). Both opinion states are typically found in the population at any given time.

\begin{figure}
    \centering
    \includegraphics[width=0.5\textwidth]{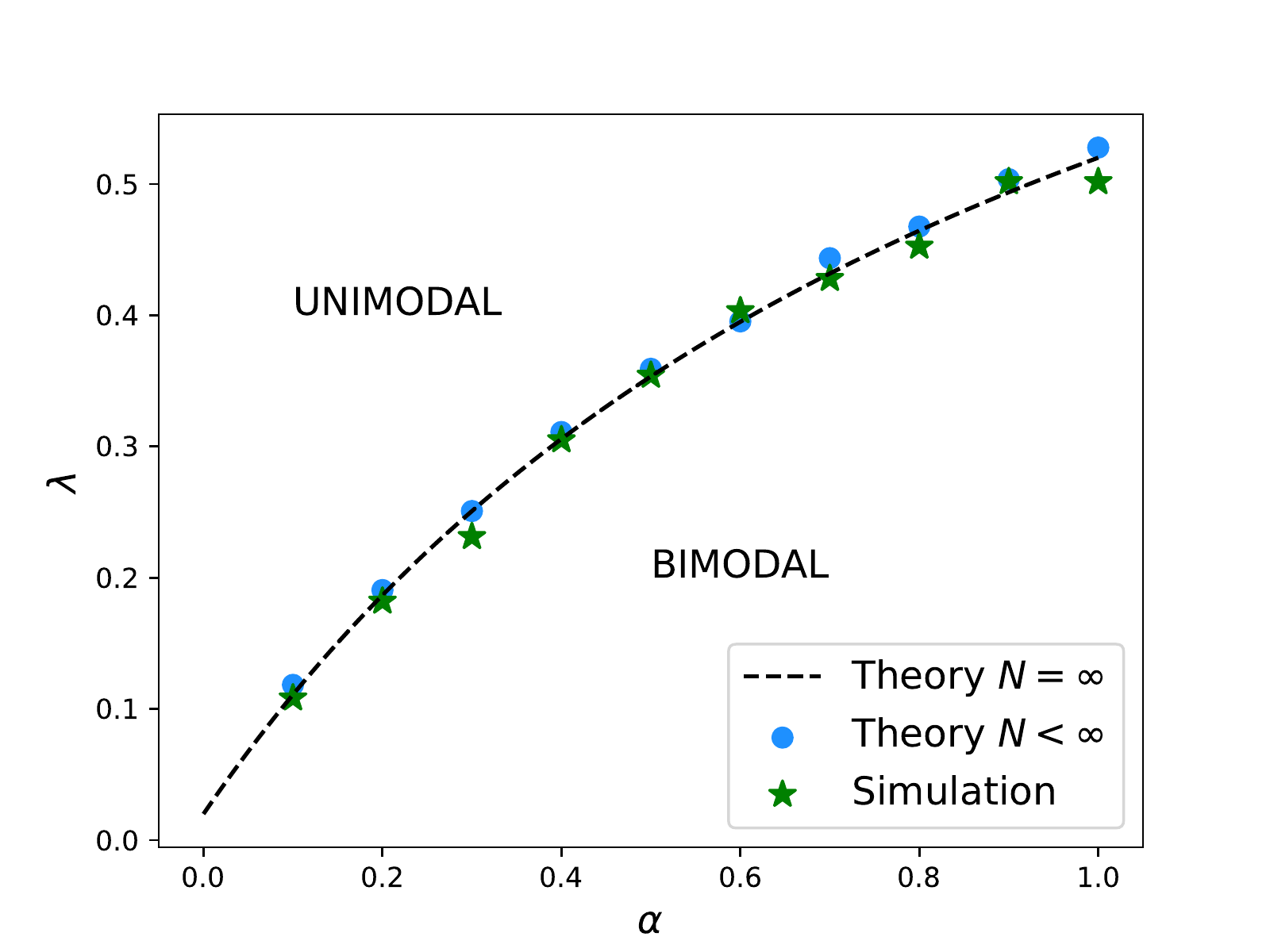}
    \caption{
     {\bf Phase diagram for the model with two  states for the group of influencers.} The dashed line is $\lambda_c$ obtained in the PDMP limit [Eq.~(\ref{eq:lambda_c_2s_pdmp})]. It separates a phase in which the stationary distribution is bimodal (below the line) from the other phase in which the distribution is unimodal (above the line). Green asterisks are from simulations of the individual-based model with $a=0.01$ and $N=500$. Blue dots indicate the phase boundary obtained from the theory which takes into account leading-order corrections to the PDMP [Eq. \eqref{eq:stat_distribution_with_noise}]. 
     Model parameters are $\overline z=1$, $\mu_{0\to 1}=\mu_{1\to 0}=1$, $a=0.01$. }
     \label{fig:Phase_diagram_2_state} 
\end{figure}

The resulting phase diagram is shown in Fig.~\ref{fig:Phase_diagram_2_state}. The system is in the unimodal state above the phase line, and in the bimodal state below the line.

\subsubsection{Lowest-order correction to the PDMP}
\begin{figure}
    \centering
    \includegraphics[width=0.44\textwidth]{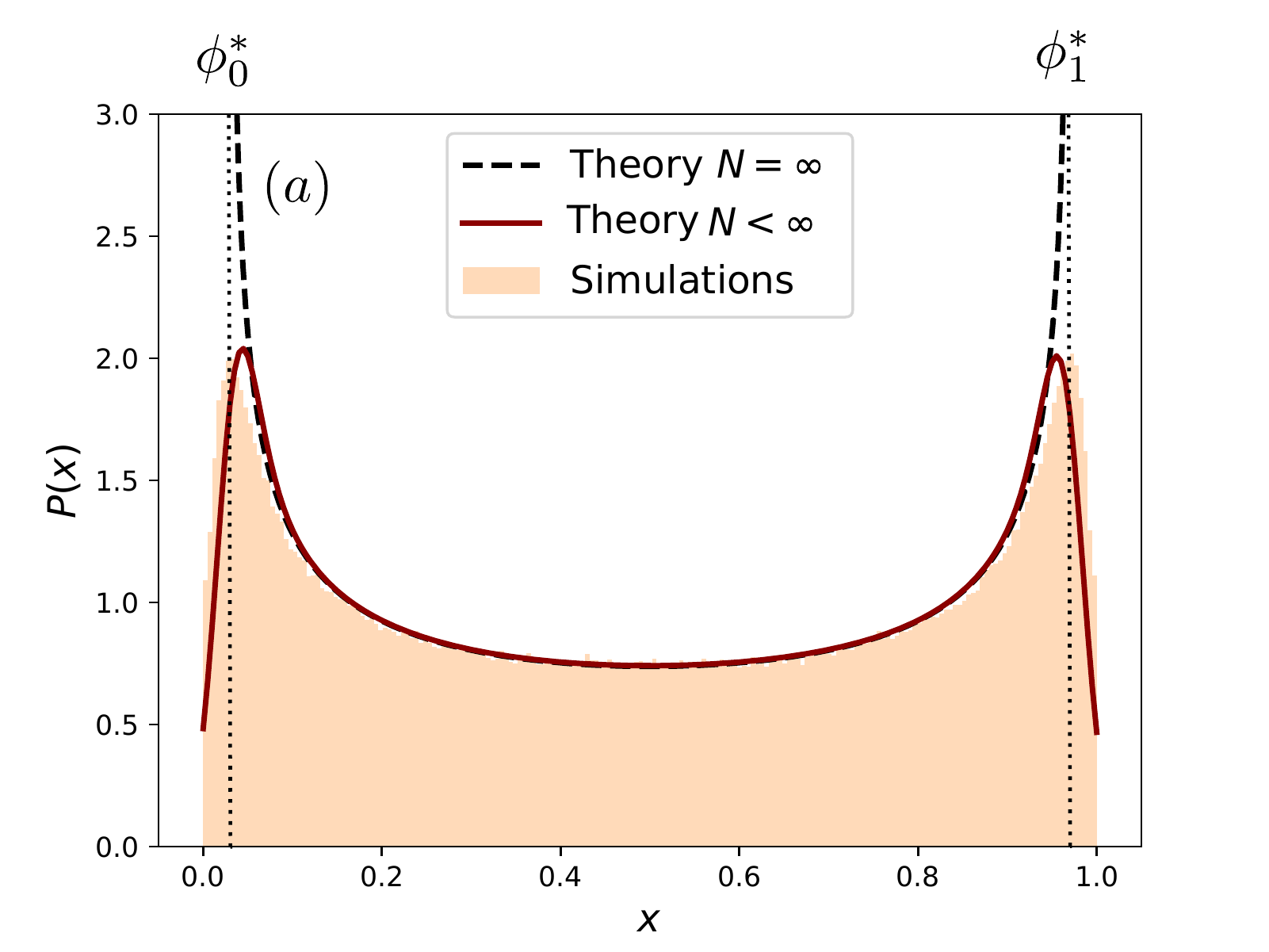}
        \includegraphics[width=0.44\textwidth]{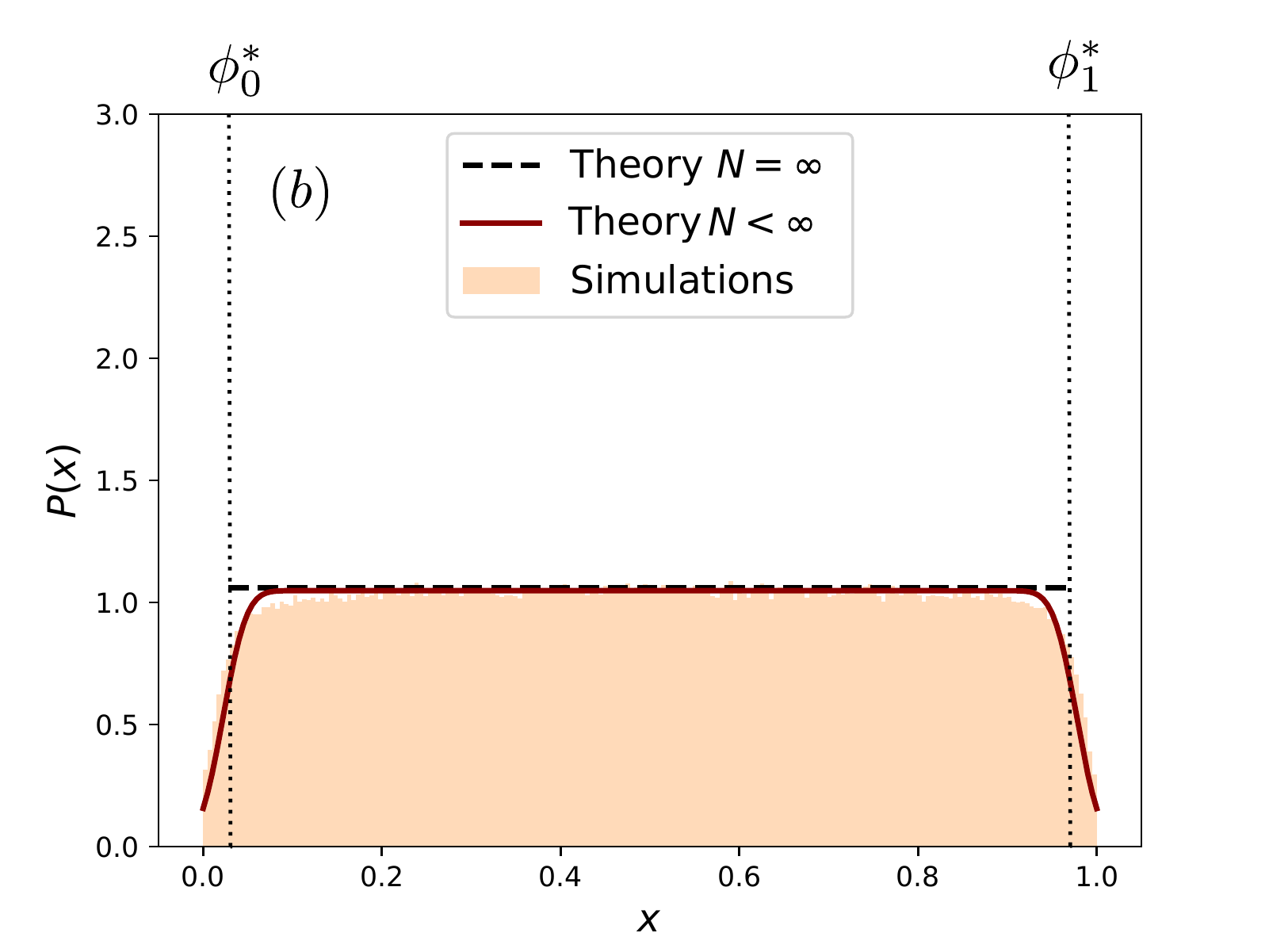}
            \includegraphics[width=0.44\textwidth]{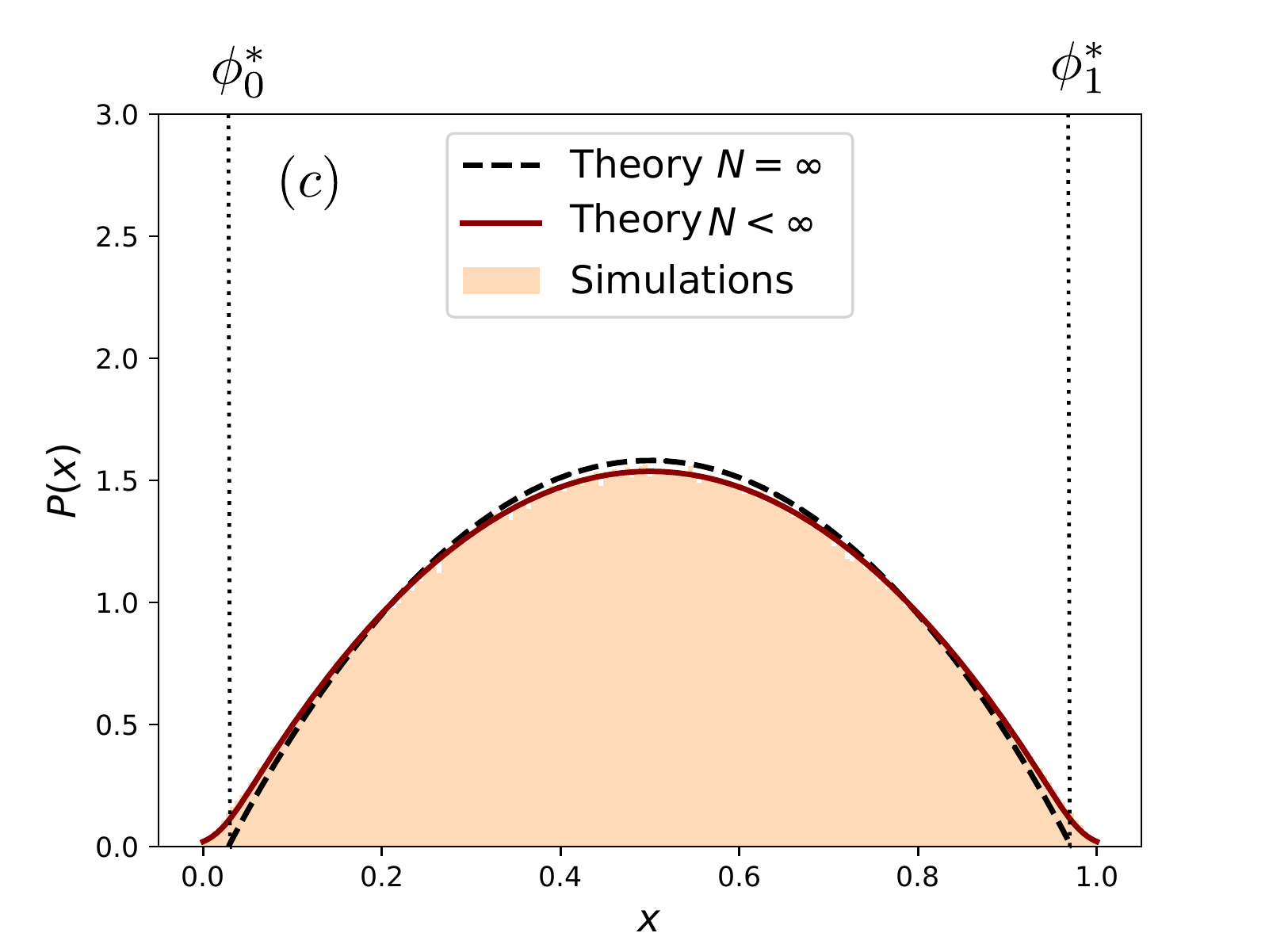}
    \caption{{\bf Stationary distribution of the model with influencers switching between two states.} The black dashed lines in each panel are from Eq.~\eqref{eq:2S_pdmp_stat} (PDMP limit), solid red lines are from the numerical integration of Eq.~ \eqref{eq:stat_distribution_with_noise}, capturing leading-order corrections to PDMP limit. The shaded histograms are from simulations of the full model. In all panels $a=0.01$, $\alpha=0.5$, $N=200$ and $\mu_{0\to 1}=\mu_{1\to 0}=1$. The switching rates are (a) $\lambda=0.2$, (b) $\lambda=\lambda_c \approx 0.35$ and (c) $\lambda=0.7$, where $\lambda_c$ is obtained from Eq.~\eqref{eq:lambda_c_2s_pdmp}. 
    Time between subsequent samples is $\Delta t=5$, for each distributions we take $10^6$ samples after a transient of $50$ units of time.}
    \label{fig:stat_distrib_2_state_N200}
\end{figure}

For the model with $\mu_\pm=1$ and $\overline z=0$ we find from Eq.~(\ref{eq:s2_1}) 
\be
s^2(\phi)=\phi(1-\phi)\left(\frac{h}{(1+\alpha)\lambda_c}+1\right).
\ee
This can be used to approximate the stationary distribution following \cite{Intrinsic}. An illustration is shown in Fig.~\ref{fig:stat_distrib_2_state_N200} where the red lines show the resulting predictions for a model with $N=200$ agents. Intrinsic noise in the model with finite populations smoothens the distribution compared to that in the PDMP limit, but the main characteristics of being bimodal or unimodal are preserved. Nevertheless, the finite size of the population results in a notable alteration in the bimodal phase. With intrinsic noise the regions $x<\phi_0^*$ and $x>\phi_1^*$ become populated. These parts of phase space are not accessible by the PDMP. Thus, intrinsic stochasticity enhances the polarization of the population.

\subsection{Three states of the group of influencers}\label{sec:3states}

In this section the group of influencers switches among three states $\sigma=0,1,2$. As before we assume that there is a state in which all influencers support opinion $B$ ($z_0=0$), and another in which all influencers favour opinion $A$ ($z_2=1$). In the intermediate state, $\sigma=1$, we assume that a fraction $\delta$ of influencers supports $A$, and a fraction $1-\delta$ acts in favour of $B$. Thus, $z_1=\delta$. Switches between these three states are taken to occur in a Markov process as follows

\be\label{eq:3states}
0 \xrightleftharpoons[\lambda/2]{\lambda }1 \xrightleftharpoons[\lambda]{\lambda/2} 2.
\ee
Thus, the environment switches out of state $0$ and to state $1$ with rate $\lambda$, and similarly for switches $2\to 1$. The total rate of leaving state $\sigma=1$ is also $\lambda$, split equally for transitions to states $0$ and $2$, respectively.

%Thus, the information on the environmental dynamics is encoded in the stochastic matrix
%\begin{equation}
%\lambda
%\left(
%\begin{array}{ccc}
%0 &  1 & 0 \\
%1/2 & 0  & 1/2 \\
%0 &  1 & 0 \\
%\end{array}
%\right).
%\end{equation}
We first discuss the model in the PDMP limit, that is for infinite populations, $N\to\infty$. The stationary state is then to be obtained from Eq.~\eqref{eq:pdmp_equation}. In the present setup this can be reduced into a system of two coupled ODE (see Appendix \ref{App:Algorithm for three state of the influencers}) but we are unable to obtain an analytical solution. However, as also explained in Appendix \ref{App:Algorithm for three state of the influencers} one can proceed numerically.

%R 

It is useful to note that the presence of three environmental states does not affect the relaxation time scale in any fixed environment. This is due to our assumption $a_\sigma\equiv a$ in all three states. Therefore, $\lambda_c$ continues to be given by the expression in Eq.~\eqref{eq:lambda_c_2s_pdmp}. 

\begin{figure}
\centering
    \includegraphics[width=0.44\textwidth]{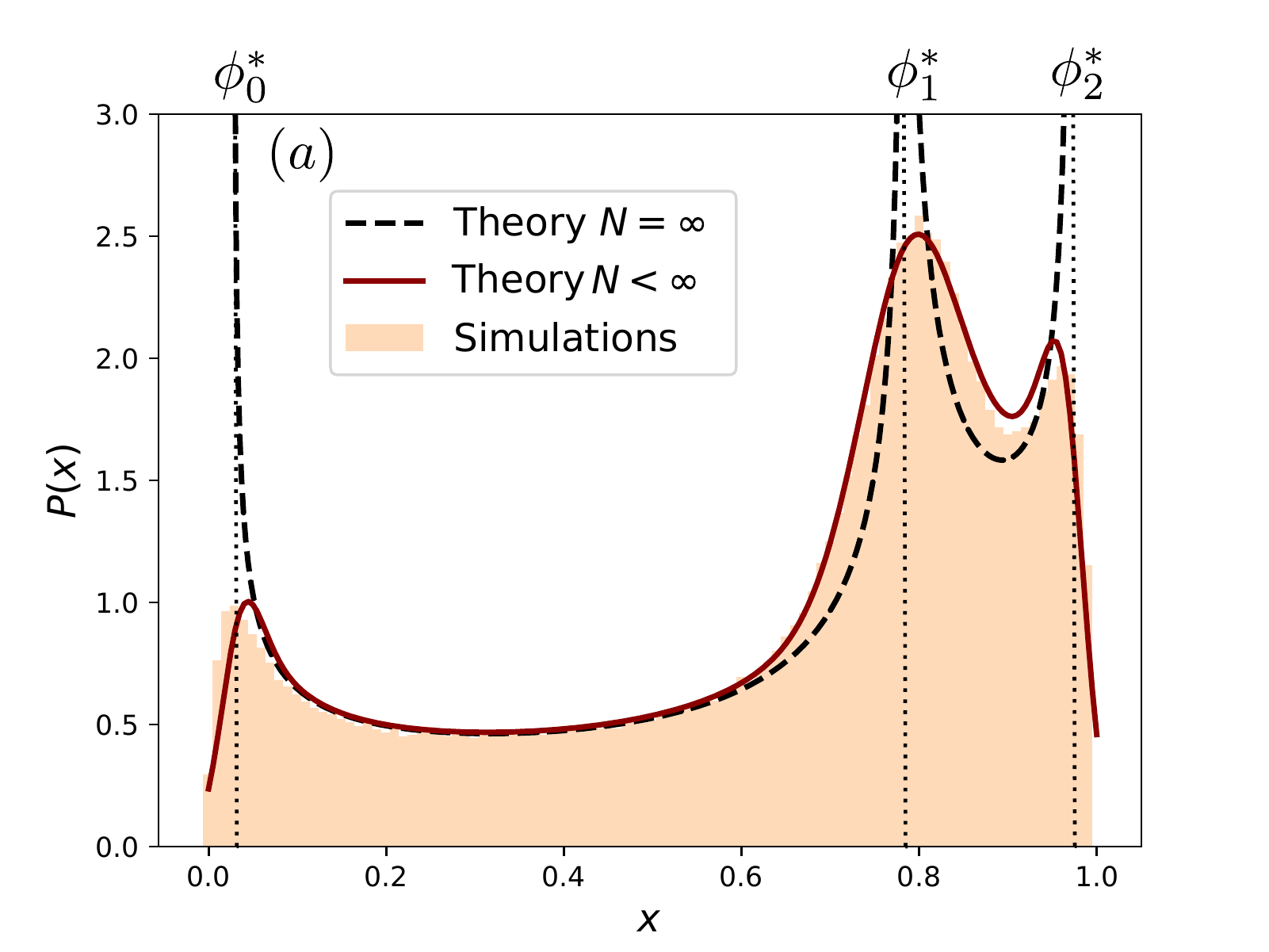}
        \includegraphics[width=0.44\textwidth]{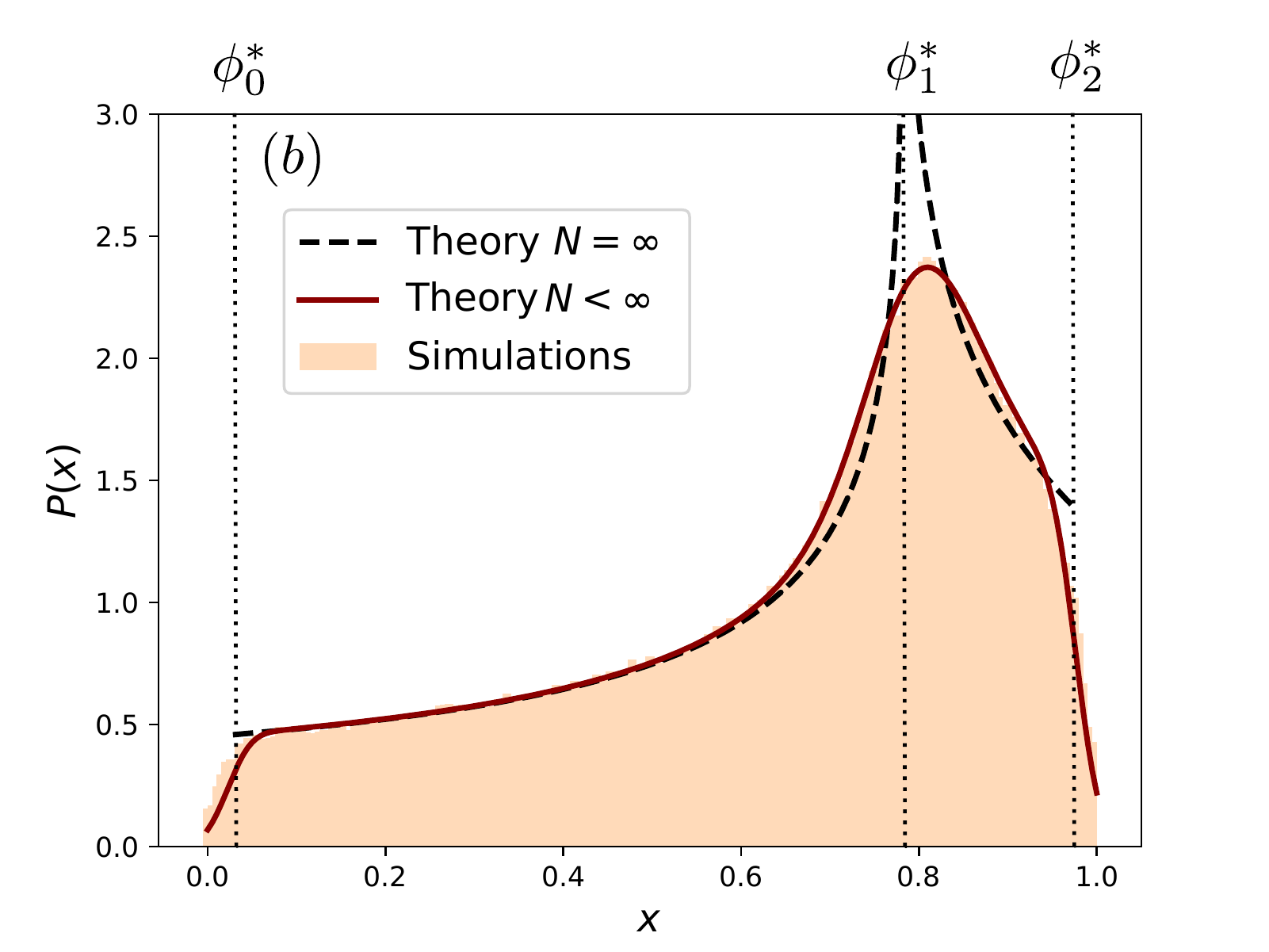}
            \includegraphics[width=0.44\textwidth]{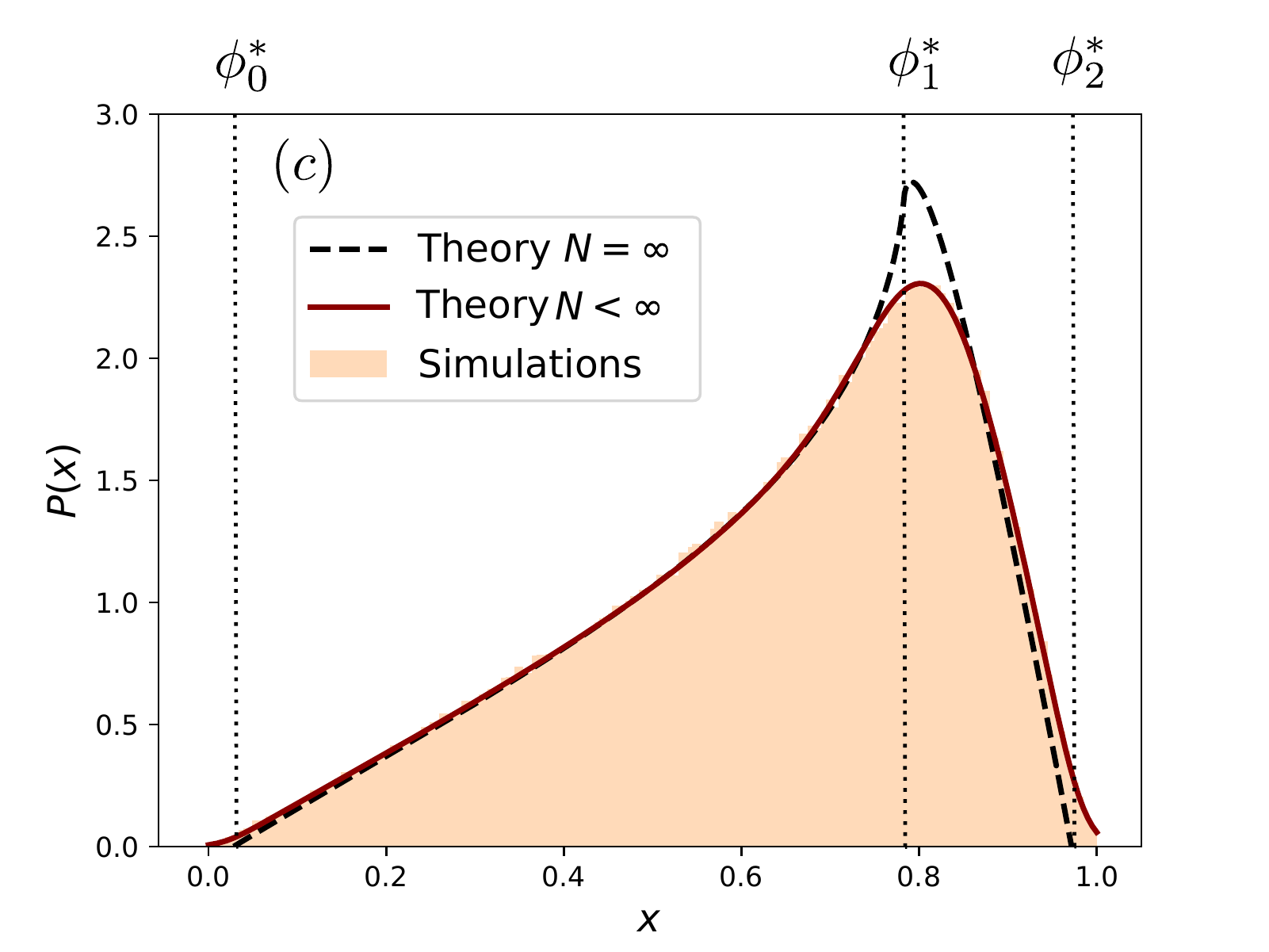}
    \caption{{\bf Stationary distributions for the model with three states for the group of influencers}. In each panel the dashed line represents the PDMP limit, and is obtained from a numerical solution of Eq.~\eqref{eq:pdmp_stat}. The solid lines are from the numerical integration of Eq.~\eqref{eq:stat_distribution_with_noise}, capturing leading-order corrections in $1/N$. The shaded histograms are from simulations of the full model. The dotted lines are the values of $\phi_0^*, \phi_1^*$ and $\phi_2^*$ found from Eq.~\eqref{eq:fixed_point}.  The environmental switching rate is $\lambda=0.2$ in panel (a), $\lambda=\lambda_c\approx 0.35$ in (b), and $\lambda=0.7$ in panel (c). In all panels $a=0.01$, $\alpha=0.5$, $N=200$, $z_0=0$, $z_1=0.8$, $z_2=1$, and $\mu_{0 \to 1}=\mu_{2 \to 1}=1$ and $\mu_{1 \to 0}=\mu_{1 \to 2}=1/2$. Time between subsequent samples is $\Delta t=5$; for each distributions we take $10^6$ samples, after a transient of $50$ units of time. }
    \label{fig:stat_distrib_3_state_N200}
\end{figure}

Results are shown in Fig.~\ref{fig:stat_distrib_3_state_N200}. We first focus on the black dashed lines showing the stationary distribution in the PDMP limit.  When environmental switching is slower than the relaxation in the population [$\lambda < \lambda_c$ shown in panel (a)] the distribution has three sharp singularities, positioned at the fixed-point values $\phi_0^*, \phi_1^*$ and $\phi_2^*$ obtained from Eq.~\eqref{eq:fixed_point} (we attribute minor numerical deviations to discretisation effects). For $\lambda >\lambda_c$ on the other hand [panel (c)], the distribution is asymmetrically unimodal with peak at $\phi_1^*$. In panel (b), where $\lambda = \lambda_c$, the distribution also has a single maximum at $\phi=\phi_1^*$. In contrast with panel (c) though, the stationary density in the PDMP limit (black dashed line) remains non-zero at $\phi=\phi^*_0$ and $\phi^*_2$ respectively. In panel (c) the density tends to zero at the bondaries.

In Fig.~\ref{fig:stat_distrib_3_state_N200} we also show results from the theory capturing the leading-order corrections in $1/N$ (solid lines). As seen intrinsic noise does not manifestly change the overall structure of the stationary distribution. Its main effect is to smoothen the singularities, and as expected there is now a non-zero probability of finding the system in the intervals $i/N\in[0,\phi_0^*]$ and $i/N\in[\phi_0^*,1]$ respectively. These intervals are (by construction) unattainable by the PDMP.

\subsection{Multiple states for groups of influencers}
\begin{figure*}
    \centering
    \includegraphics[width=0.48\textwidth]{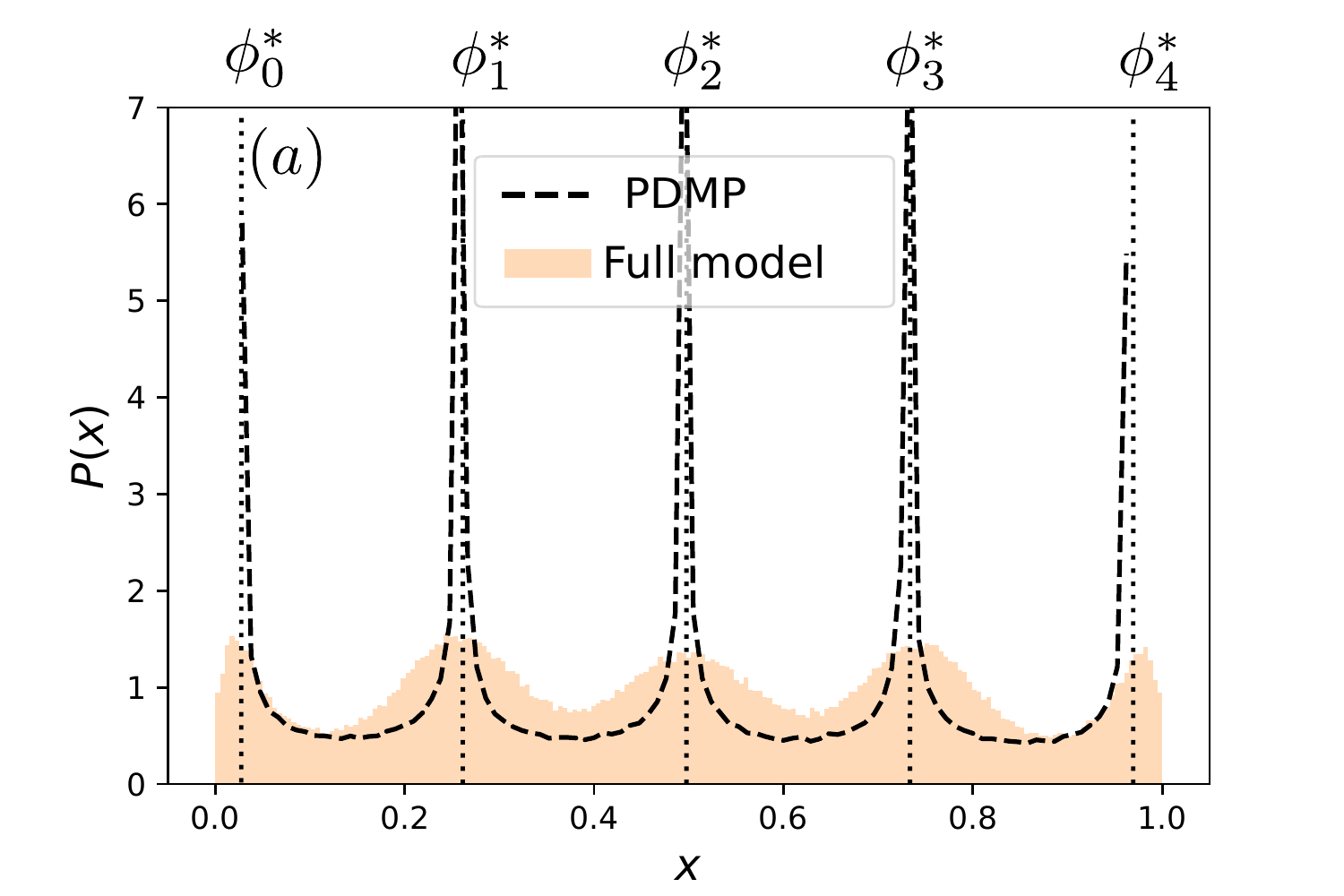}
    \includegraphics[width=0.48\textwidth]{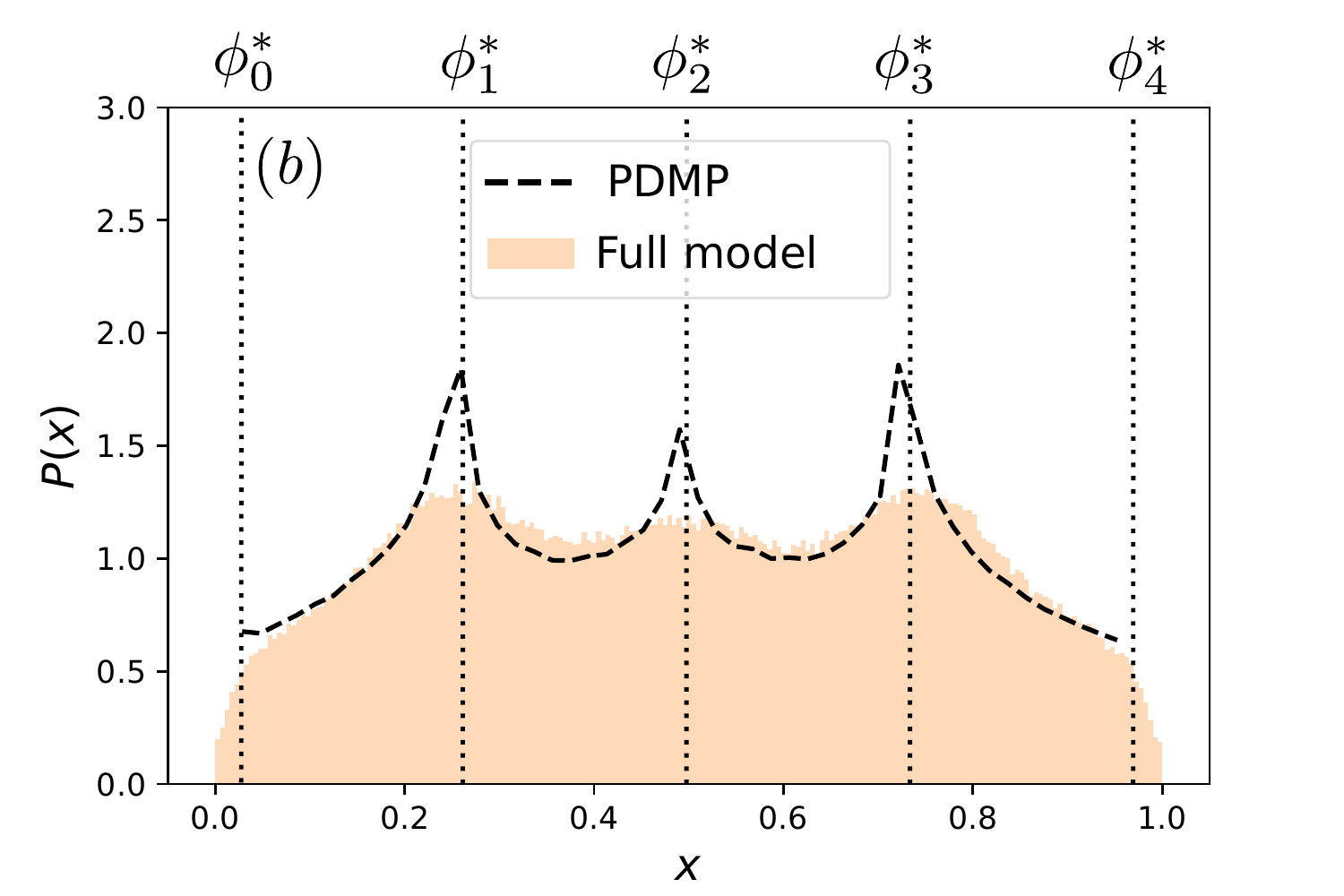}
    \includegraphics[width=0.48\textwidth]{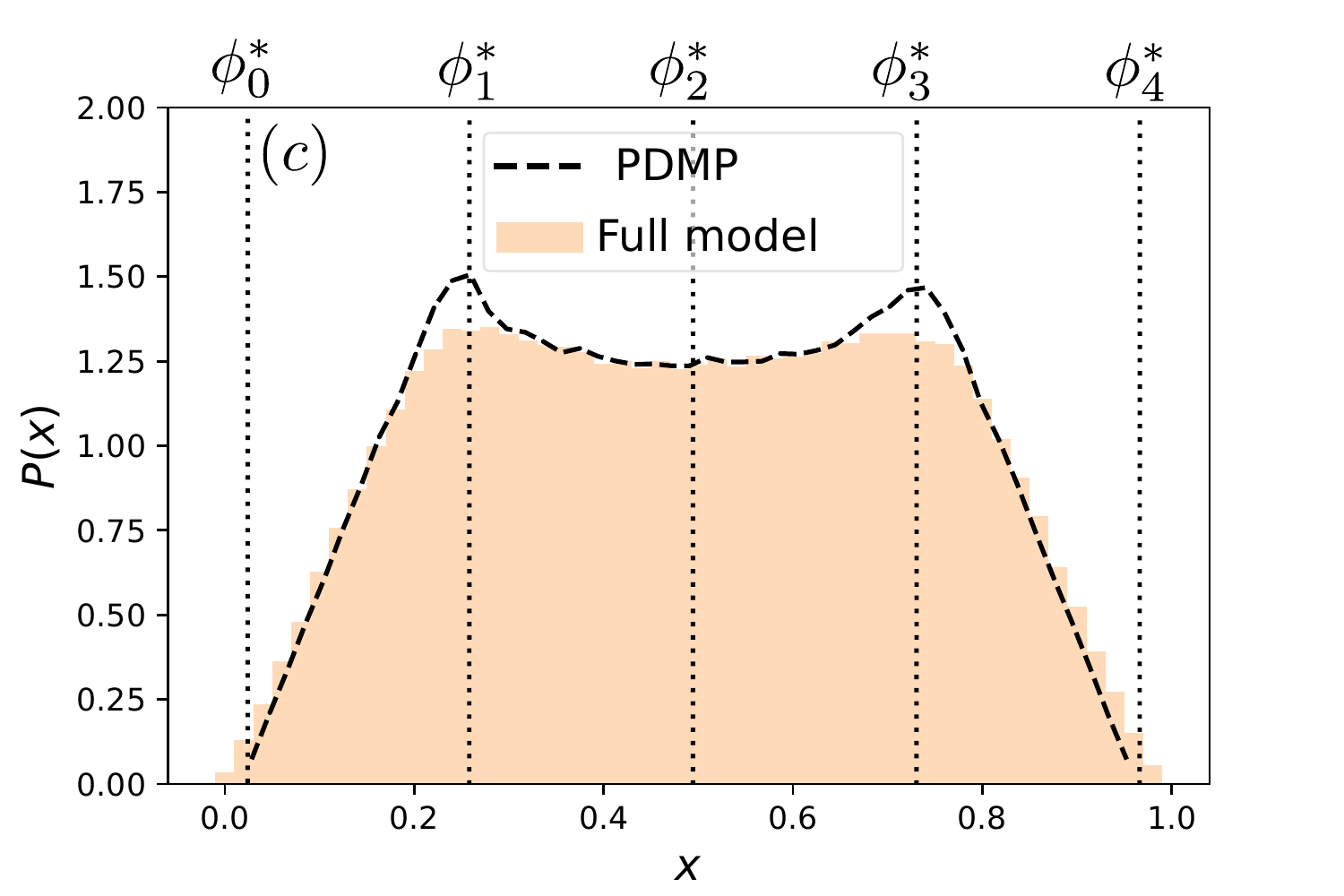}
    \includegraphics[width=0.48\textwidth]{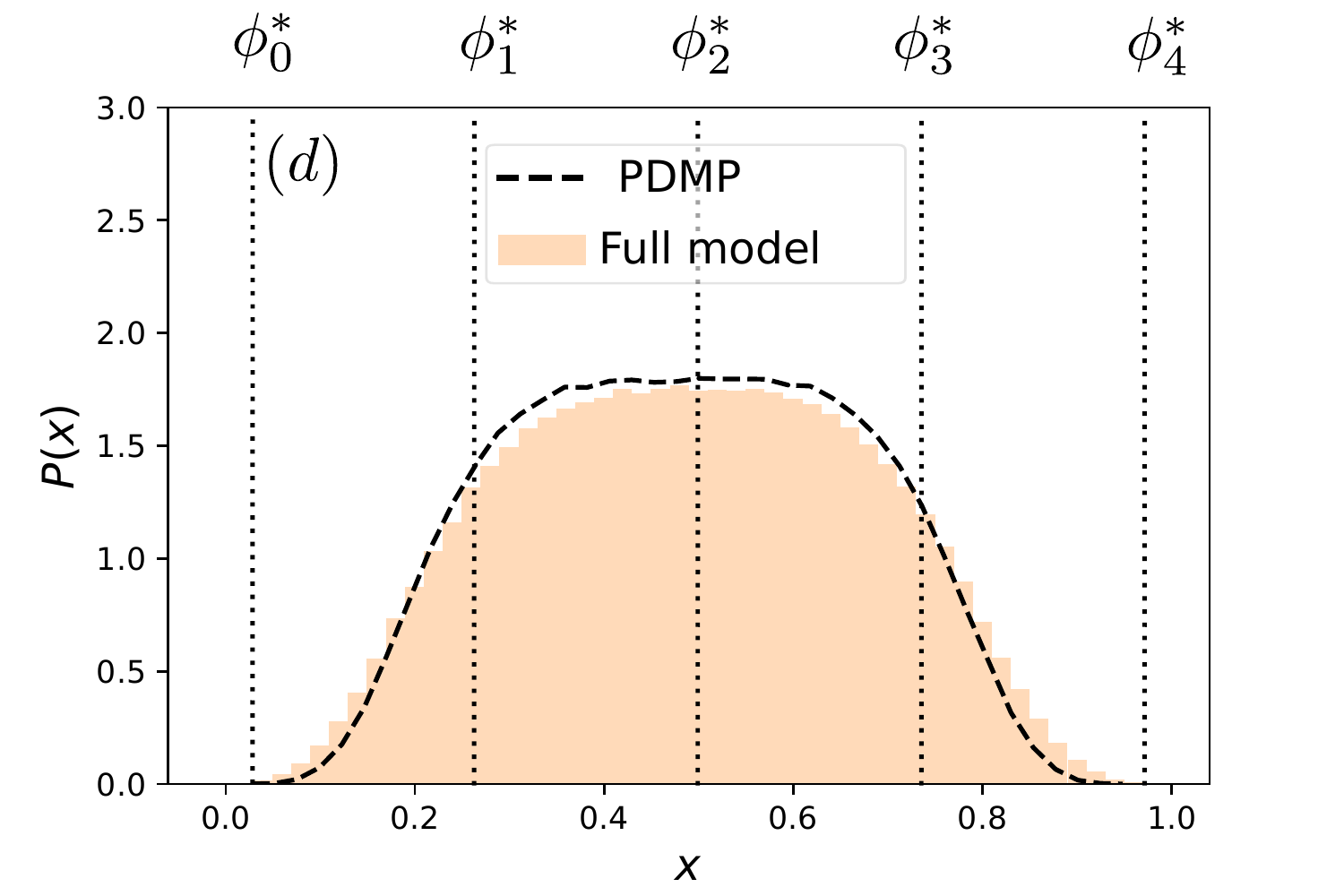}
    \caption{{\bf Stationary distribution for the model with five environmental states.} The dashed lines in each panel are from numerical simulations of PDMP capturing the limit of infinite populations. Shaded histograms are from simulations of the full individual-based model with $N=200$. The environmental dynamics is as in Eq.~(\ref{eq:5states}). The switching rates in panels (a)-(d) are $\lambda=0.1$, $\lambda=\lambda_c\approx 0.35$, $\lambda=0.7$ and $\lambda=2$ respectively.
    Time between samples is $\Delta t=1$ in (a)-(c), and $\Delta t=0.2$ for panel (d). We take $10^7$ samples after a transient of $10$ units of time.}
    \label{fig:stat_distrib_manyS_N200}
\end{figure*}
We now focus on systems in which there are more than three states for the environment of influencers. The numerical solution of Eq.~\eqref{eq:pdmp_equation} then becomes more complex, and we hence focus on direct simulations of the original individual-based model, and of the limiting PDMP respectively.

\subsubsection{Five environmental states}\label{sec:5states}
We first focus on a generalisation of the system in Eq.\eqref{eq:3states} to five environmental states,
\be\label{eq:5states}
0 \xrightleftharpoons[\lambda/2]{\lambda }1 \xrightleftharpoons[\lambda/2]{\lambda/2} 2 \xrightleftharpoons[\lambda/2]{\lambda/2} 3
\xrightleftharpoons[\lambda]{\lambda/2} 4.
\ee
We set $z_\sigma=\sigma/4$ for $\sigma=0,1,\dots,4$. Thus in state $\sigma=0$ all influencers promote opinion $B$, and for $\sigma=4$ the external force is fully in direction of opinion $A$. State $1$ is partially biased towards $B$, in state $\sigma=2$ there is no net force by the influencers in either direction, and $\sigma=3$ represents a state with partial bias towards opinion $A$.

In Fig.~\ref{fig:stat_distrib_manyS_N200}
we show stationary distributions from simulations of the full model for $N=200$ and with different choices of the switching rate $\lambda$ (shaded histograms). We also show the stationary distributions from simulations of the PDMP (dashed lines). 

In panel (a) we choose $\lambda<\lambda_c$, i.e., the population relaxes faster than the time between switches of the environment. We observe five singularities in the stationary distribution of the PDMP, located at the different $\phi_\sigma^*$. As before, intrinsic noise smoothens these peaks. 
Panel (b) shows the case $\lambda=\lambda_c$, we then find three peaks in the stationary distribution of the PDMP. These maxima are also discernible in the stationary distribution of the full model, but the intrinsic noise smears the distribution out, so that the maxima are less pronounced. Increasing the rate of influencer switching further [panel (c)], the number of maxima reduces to two, and finally in panel (d) the stationary state becomes unimodal.

\begin{figure}
    \centering
    \includegraphics[width=0.48\textwidth]{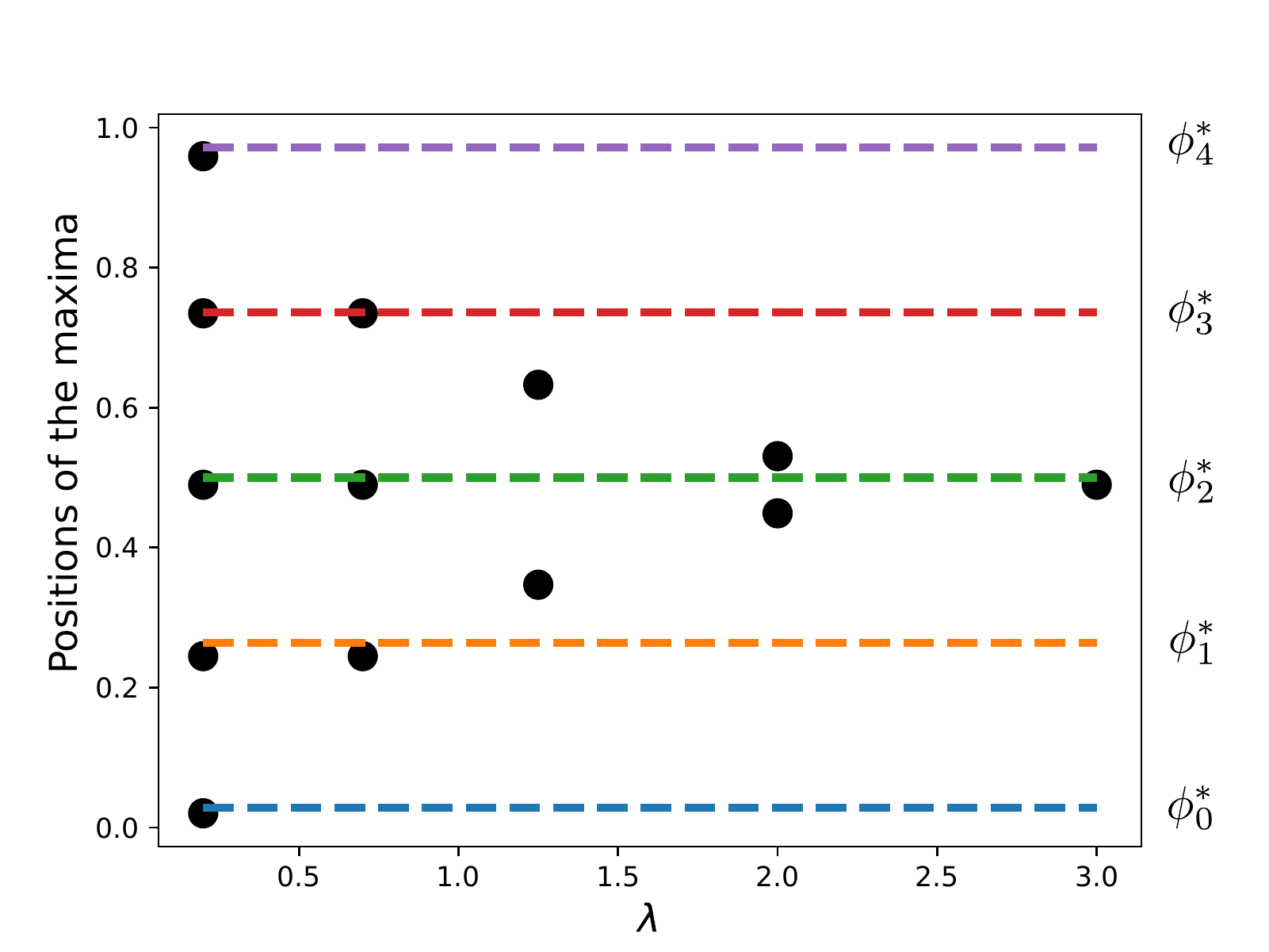}
          \caption{{\bf Location of the maxima of the stationary distribution for the model with five environmental states (Fig.~\ref{fig:stat_distrib_manyS_N200}).}The figure shows the location of maxima in the stationary distribution of the PDMP for the model with five environmental states (Sec.~\ref{sec:5states}). The markers are from simulations of the PDMP, the lines indicate the fixed points $\phi_0^*,\dots,\phi_4^*$. Except for $\lambda$ parameters are as in Fig.~\ref{fig:stat_distrib_manyS_N200}.}
    \label{fig:maxima}
\end{figure}

The positions of the maxima are shown in Fig.~\ref{fig:maxima} for different values of the switching rate $\lambda$. For small $\lambda$ there are five maxima, located at the $\phi_\sigma^*$. For intermediate switching rates, only three maxima remain, located at their initial positions $\phi_1^*, \phi_2^*, \phi_3^*$. Next the maximum at $\phi_2^*$ disappears. Finally, the transition to only only maximum at large values of $\lambda$ on the other hand occurs by gradual approach and eventual fusion of the two remaining maxima.

\begin{figure*}
    \centering
    \hspace{0.4cm}\includegraphics[width=0.48\textwidth]{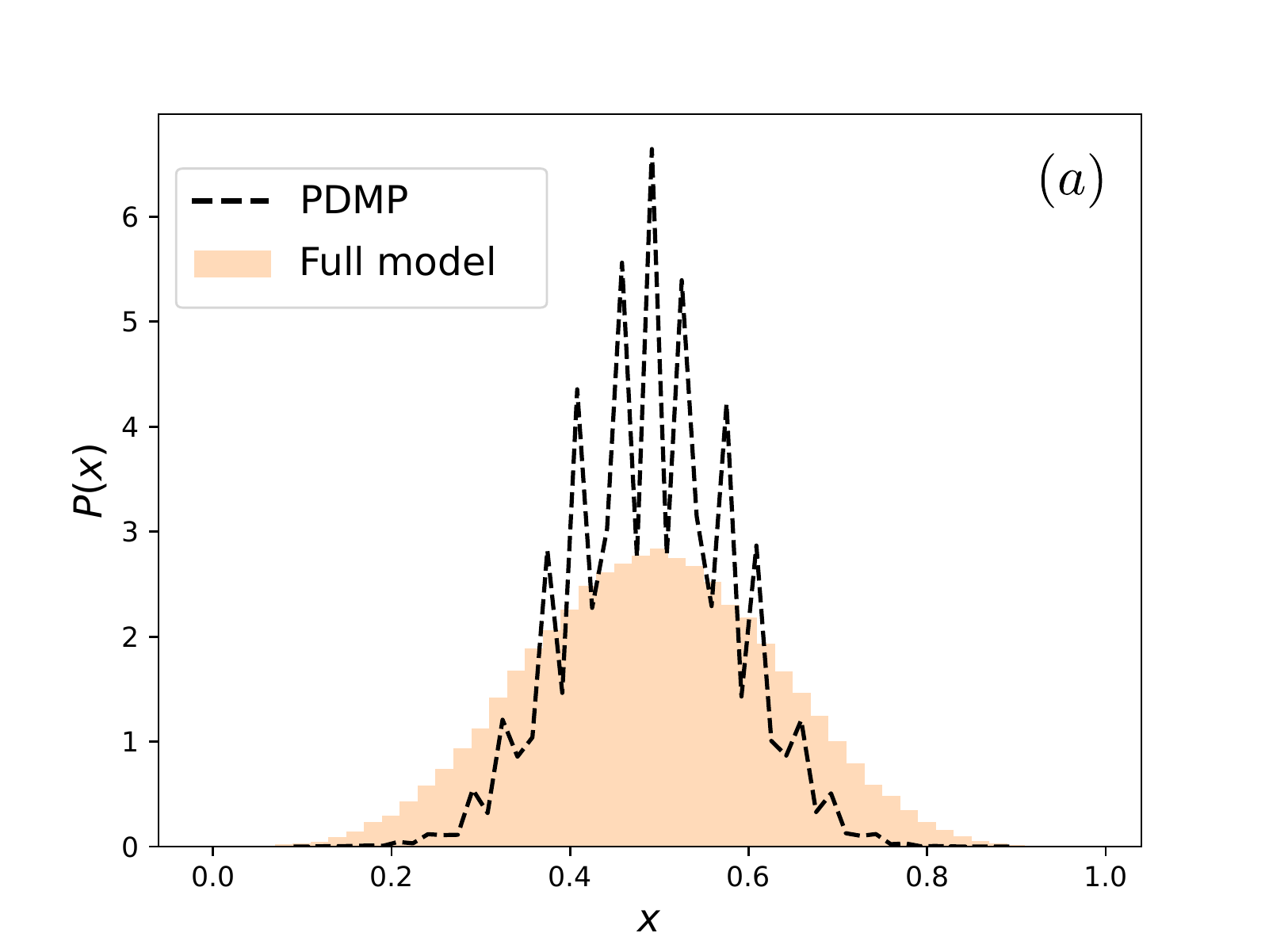}
    \includegraphics[width=0.48\textwidth]{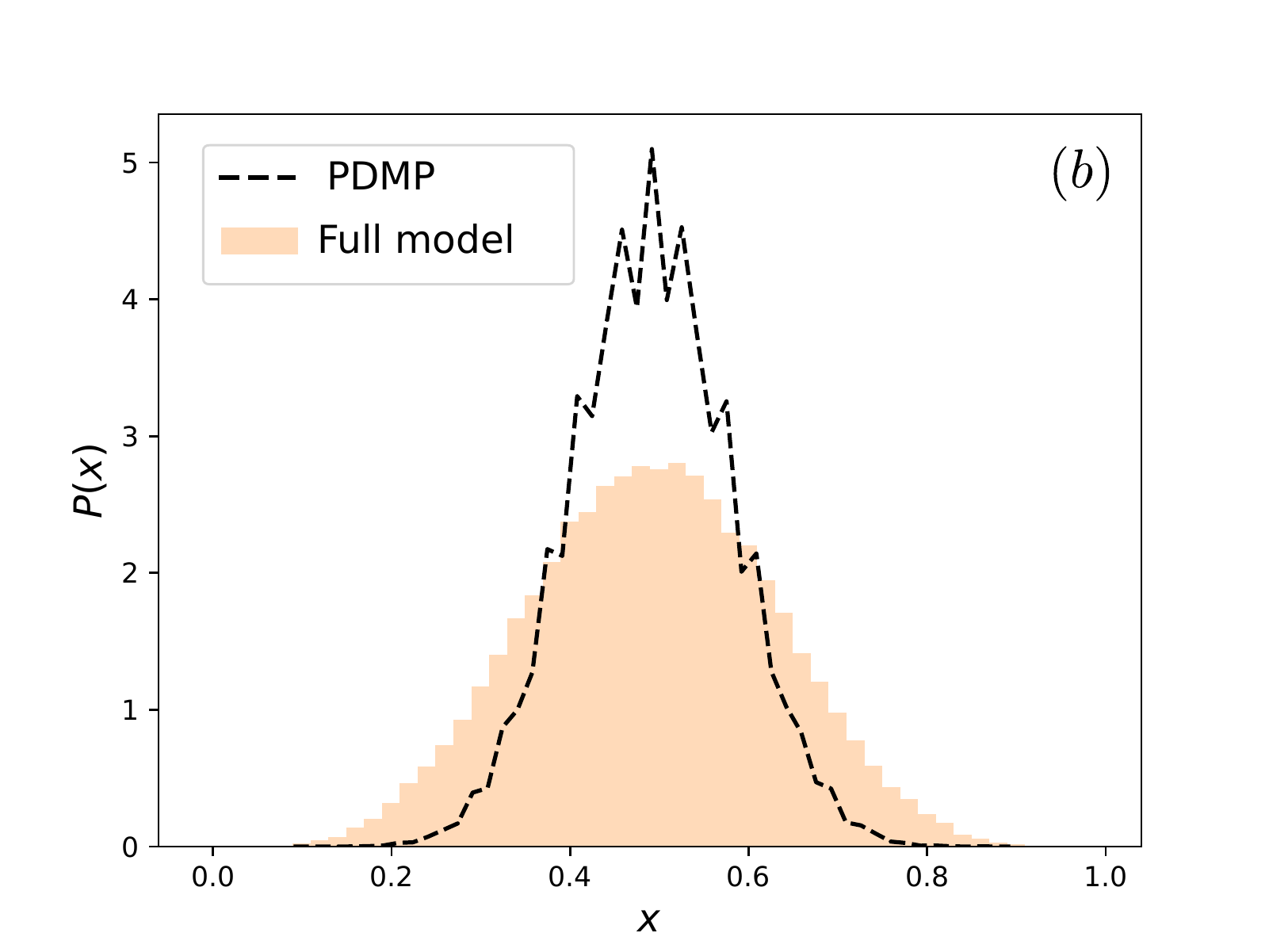}
    \includegraphics[width=0.48\textwidth]{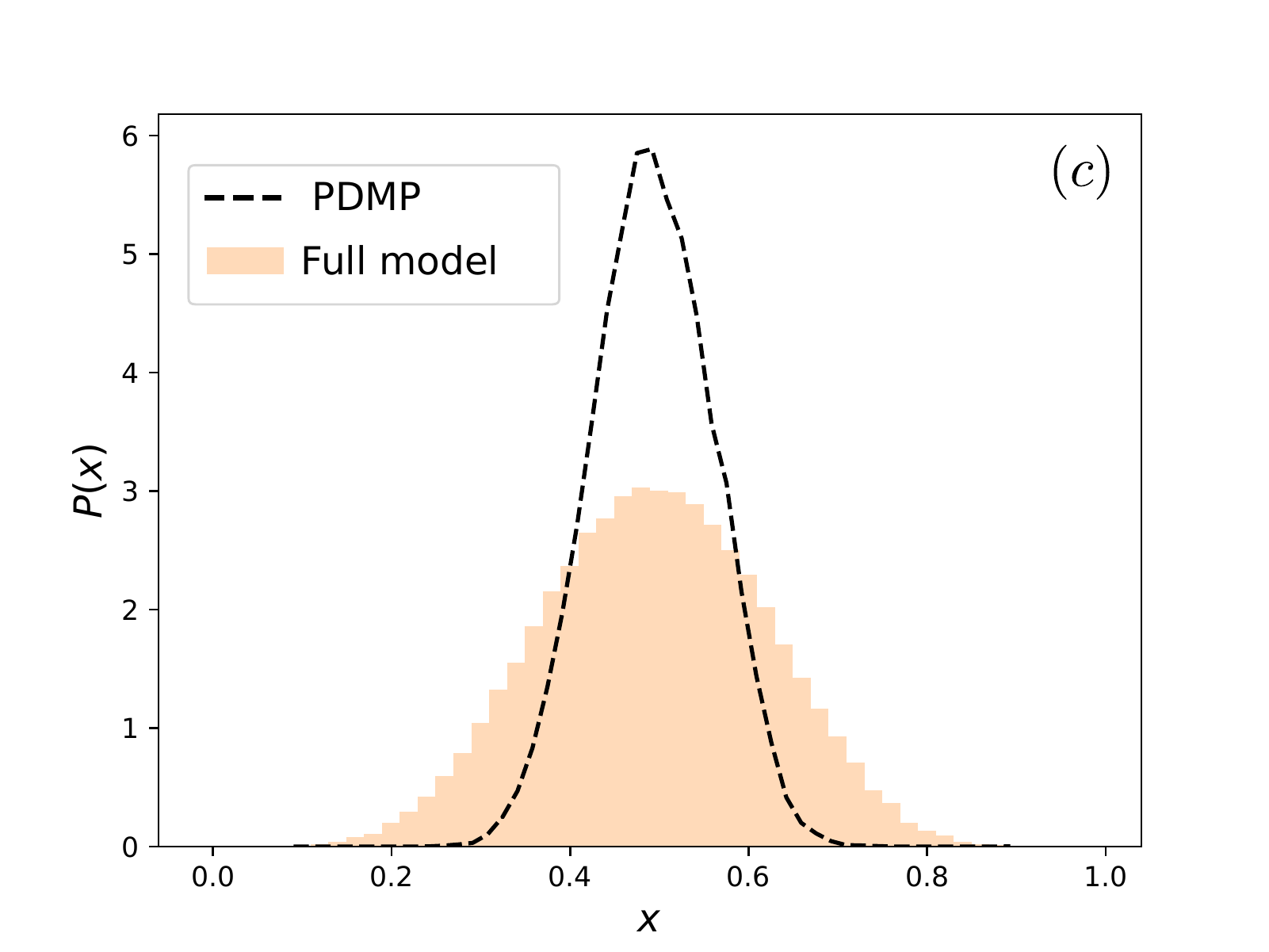}
    \includegraphics[width=0.45\textwidth]{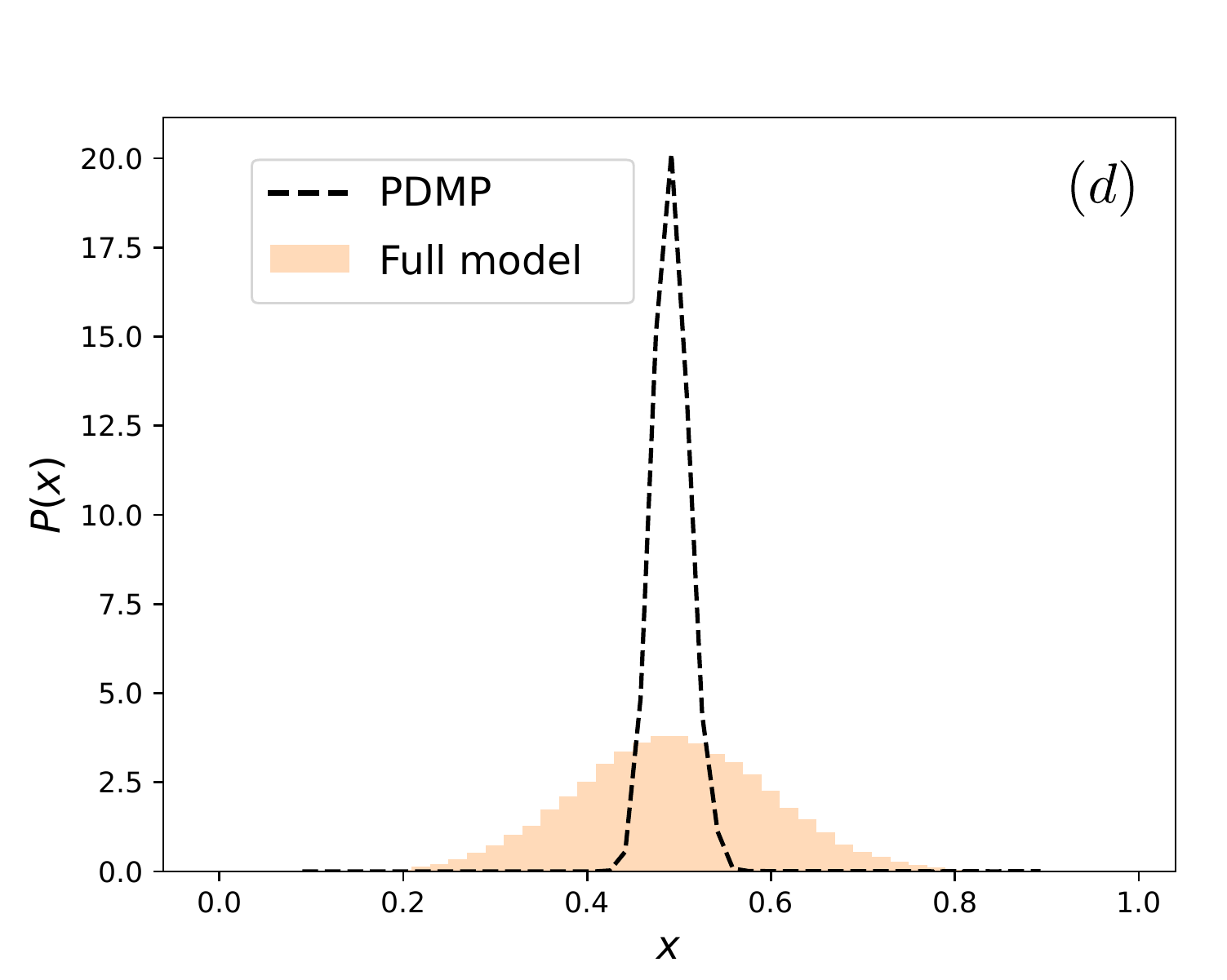}
    \caption{ \textbf{Stationary distribution for the model with independent influencers.}
    The influencers follow the process in Eq.~\eqref{eq:independent_influencers}. We use $z_\sigma=\sigma/(\alpha N)$ with $\sigma=0,...,\alpha N$, with $\alpha N=20$. Dashed lines in each panel are from numerical simulations of the PDMP, the shaded histogram is from simulations ($N=200$). Model parameters are $a=0.01$, $\alpha=0.1$. Environmental switching rate is (a) $\lambda=0.05$, (b) $\lambda_c\approx 0.11$, (c) $\lambda=1$ and (d) $\lambda=20$. 
    Time between subsequent samples is $20$ units of time in (a) and (b), $2$ units in (c), and $0.2$ units of time in (d). For each distributions we take $10^7$ samples after a transient of duration $100/\lambda$. }
    \label{fig:IndependentInflu_2}
\end{figure*}
\subsubsection{Independent influencers}

Next, we consider the case of independent influencers. The influencers are all taken to have the same strength, and each influencer can act in favour of opinion $A$, or of opinion $B$. In the example in Fig.~\ref{fig:IndependentInflu_2} there are $\alpha N=20$ influencers, and hence $\alpha N+1$ environmental states $\sigma=0,1,\dots, S-1=\alpha N$. In state $\sigma$ there are $\sigma$ influencers favouring $A$, and $S-1-\sigma$ influencers promoting $B$. Thus, $z_\sigma=\sigma/(S-1)$. 

In state $\sigma$ there are $\sigma$ influencers who can switch to promoting $B$ instead of $A$, and $S-1-\sigma$ influencers who can change from favouring $B$ to favouring $A$. Thus, the rate of transitioning from state $\sigma$ to state $\sigma-1$ is proportional to $\sigma$, and that of transitioning from $\sigma$ to $\sigma+1$ is proportional to $S-1-\sigma$. We set $\mu_{\sigma\to\sigma-1}=\sigma/(S-1)$ and $\mu_{\sigma\to\sigma+1}=1-\sigma/(S-1)$. Keeping in mind the overall multiplying factor $\lambda$, the environmental dynamics can then be summarised as
\be\label{eq:independent_influencers}
0 \xrightleftharpoons[\lambda z_1]{\lambda (1-z_0)}
%1  \xrightleftharpoons[\lambda z_2]{\lambda(1-z_1)} 
\dots
\xrightleftharpoons[\lambda z_\sigma]{\lambda(1-z_{\sigma-1})} \sigma
\xrightleftharpoons[\lambda z_{\sigma+1}]{\lambda (1-z_\sigma)} \dots
\xrightleftharpoons[\lambda z_{S-1}]{\lambda(1-z_{S-2})} S-1.
\ee
Effectively, this means that each one of the $\alpha N$ individual influencers changes state with rate $\lambda/(S-1)$. We note that the division by $S-1$ is immaterial as any constant factors can be absorbed into the overall multiplier $\lambda$.

The resulting stationary distributions in Fig.~\ref{fig:IndependentInflu_2} show some of the behaviour seen in the previous example in Fig.~\ref{fig:stat_distrib_manyS_N200}. For slow environmental switching the distribution has multiple maxima in the PDMP approximation. The number of extrema decreases with increasing switching rate of the environment, and ultimately only one single maximum remains [panels (c) and (d)]. Carrying out simulations of the full model for a population of the same size ($N=200$) as in Fig.~\ref{fig:IndependentInflu_2} we find in Fig.~\ref{fig:stat_distrib_manyS_N200} that the stationary distribution is unimodal throughout. This is a consequence of the fact that the maxima of the PDMP for slow switching [Fig.~\ref{fig:stat_distrib_manyS_N200}(a)] are found relatively closely to each other. Intrinsic noise therefore `washes out' this structure much more easily than in  Fig.~\ref{fig:IndependentInflu_2}(a), where the maxima for the PDMP are more separated.

\subsubsection{Details of the influencer dynamics matter}
\begin{figure*}
    \centering
    \includegraphics[width=0.45\textwidth]{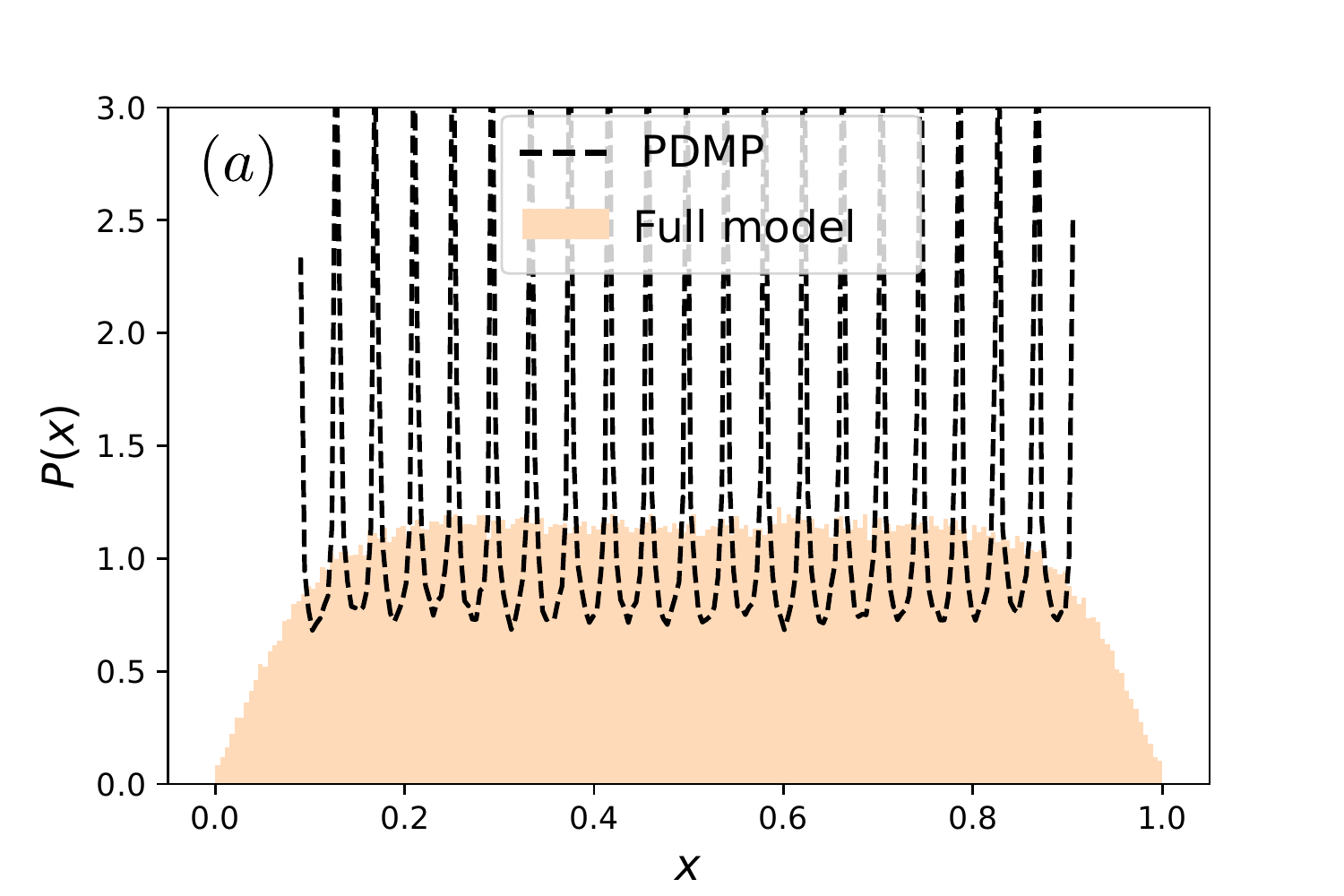}        \includegraphics[width=0.45\textwidth]{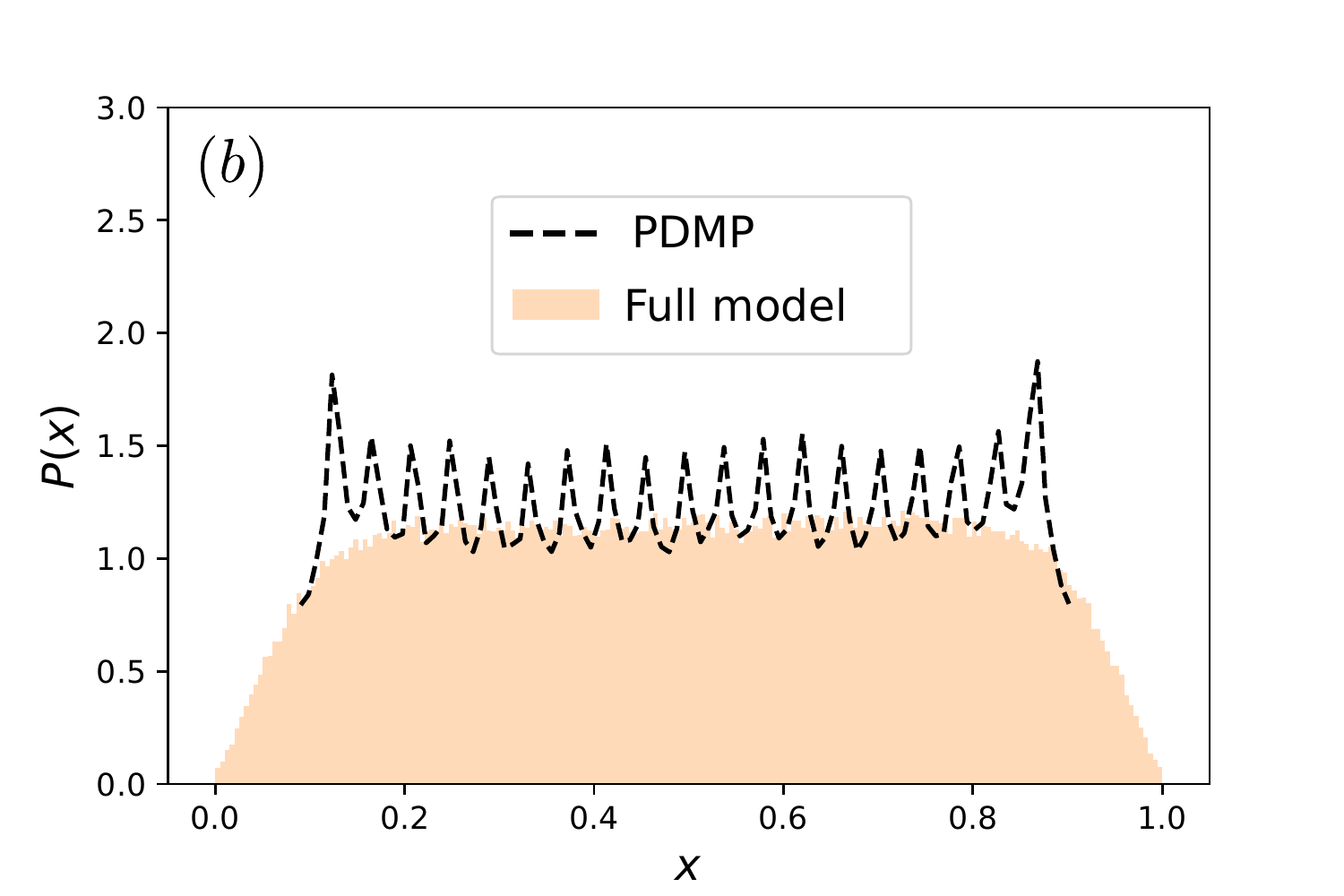}
    \includegraphics[width=0.45\textwidth]{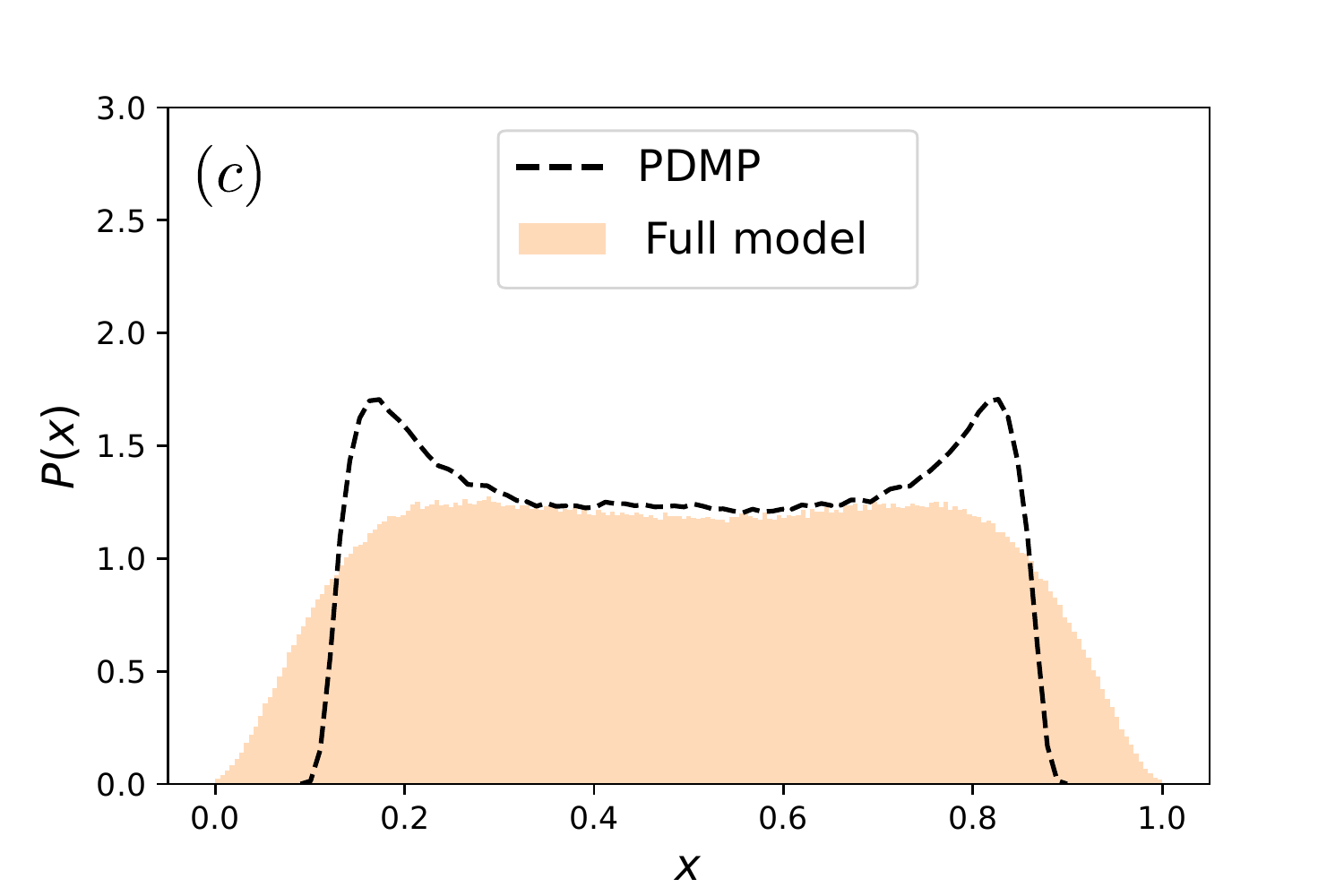}
    \includegraphics[width=0.45\textwidth]{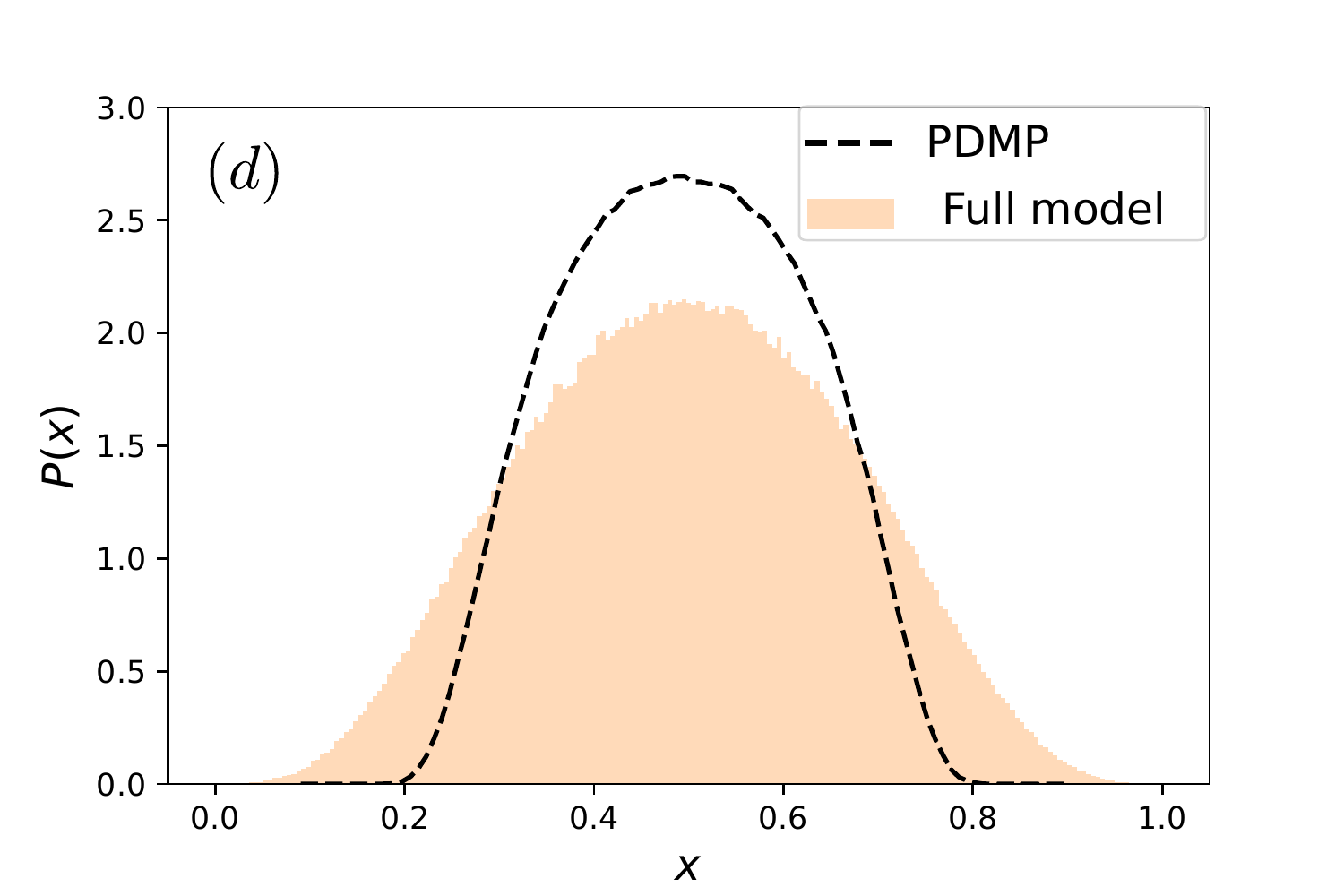}
    \caption{{\bf Stationary distribution for the model with environmental dynamics as in Eq.~(\ref{eq:env20states}).} Dashed lines are from numerical simulations of PDMP process, shaded histograms are from simulations of the full model ($N=200$). In all the panels $a=0.01$, $\alpha=0.1$. The environmental switching rates are (a) $\lambda=0.05$, (b) $\lambda=\lambda_c \approx 0.11$, (c) $\lambda=1$ and (d) $\lambda=20$. We use $z_\sigma=\sigma/(\alpha N)$ with $\sigma=0,...,\alpha N$; where $ \alpha N =20$.
    Time between subsequent samples $20$ units of time in (a) and (b), two units of time in (c), and $0.2$ units of time in (d). For each distribution we take $10^7$ samples after a transient of length $100/\lambda$.}
    \label{fig:20states}
\end{figure*}

To characterise the relation between the distributions in Figs.~\ref{fig:IndependentInflu_2} and ~\ref{fig:stat_distrib_manyS_N200} further, we study an intermediate scenario. As in Fig.~\ref{fig:stat_distrib_manyS_N200} we allow for $\alpha N+1=21$ states of the environment, and we use $z_\sigma=\sigma/(S-1)$. However, influencers no longer switch states independently from one another, but instead the environmental state is governed by a process more akin to that in Eq.~(\ref{eq:5states}). Specifically, we focus on
\begin{equation}\label{eq:env20states}
\hspace{-1cm} 0 \xrightleftharpoons[\lambda/2]{\lambda }1 \xrightleftharpoons[\lambda/2]{\lambda/2} \text{...}\xrightleftharpoons[\lambda/2]{\lambda/2} \alpha N-1
\xrightleftharpoons[\lambda]{\lambda/2} \alpha N \end{equation}

The stationary distribution for the model with independent influencers  [Fig.~\ref{fig:IndependentInflu_2}] was found to be unimodal throughout for the parameters we tested, and the corresponding PDMP has a unimodal envelope, modulated by local maxima for slow switching. 

In contrast, if the environmental dynamics is as given in Eq.~(\ref{eq:env20states}) the envelope of the stationary distribution of the PDMP is more flat at least for slow and intermediate environmental switching rates [Fig.~\ref{fig:20states} (a)-(c)]. We again find a modulation and the resulting maxima. The stationary distribution of the full model (i.e., including intrinsic noise) also has a broader shape than that in Fig.~\ref{fig:IndependentInflu_2}.

These differences in outcome can only be attributed to the differences in the environmental process [Eq.~(\ref{eq:independent_influencers}) vs Eq.~(\ref{eq:env20states})]. In the former case the environment has a proclivity to move from the more extreme states (those near $\sigma=0$ and $\sigma=\alpha N$ respectively) towards the more balanced states (those with values of $\sigma$ close to $\alpha N/2$). As a consequence the stationary distribution of the environmental process will be concentrated on the balanced states. In contrast, the dynamics in Eq.~(\ref{eq:env20states}) results in a lower tendency for the influencers to populate the more balanced states. As result of that, in turn, the stationary distribution for the population of voters becomes more broad.

\section{Conclusions}\label{sec:Conclusion}

In summary we have analysed variants of the noisy voter model with two opinion states subject to an external environment, which switches between discrete states following a random process. Specifically, we considered a switching ratio of herding-to-noise rates, and, separately, fluctuating external groups of `influencers' acting on the population of voters.

We find that the model with switching herding-to noise ratio can be reduced to a standard nVM in the limit of very fast environments. One then observes the familiar finite-size transition between a unimodal stationary state for large populations, and a bimodal state for small populations. When the environmental process is much slower than the relaxation time scale of the voters an additional trimodal phase is found for intermediate population sizes. There are then periods in which the population of voters is polarised (this occurs when herding is strong). At other times (when herding is weak) both opinions co-exist.

When influencers switch between two symmetric states (at constant herding and noise rates) we also find a transition between unimodal and bimodal states. In the limit of fast influencers the resulting phase diagram [Fig.~\ref{fig:f_s_2_state_sym}~(b)] can again be understood via a mapping to a conventional nVM with an effective noise rate. For very large populations the transition can alternatively be studied in terms of a piecewise deterministic Markov process and corrections to it. The transition between unimodal and bimodal phases can then for example be observed as a function of the strength of influencers and the environmental switching rate [Fig.~\ref{fig:Phase_diagram_2_state}]. If the two states of the influencers are not symmetric, we find an additional phase in which the stationary distribution is monotonic [Fig.~\ref{fig:s_2_state_asym}].

If there are more than two states for the external influencers the complexity of the stationary distribution of opinions also increases. For large populations (PDMP limit) we find stationary states with multiple sharp peaks when the influencer switching is slow. For higher switching rates the number of maxima generally reduces, and for very fast switching only a single peak remains, corresponding to coexistence of the two opinions. Intrinsic noise in finite populations washes out the sharp peaks seen for the PDMP, but the general trend tends to remain, there are multiple peaks for the distribution of opinions when the environment is slow, and gradually fewer peaks as influencers change states more often. We have demonstrated that the precise shape of the resulting stationary state and the location of the peaks depend on the detailed mechanics of the influencer process.

Our work thus contributes to a research programme of continuously extending the basic mechanics of the voter model. In particular, it is aligned with other recent work on variants of the voter model with fluctuating environments  \cite{KUDTARKAR,mobilia2023polarization}. While the basic voter model can be understood as a crude and stylised characterisation of opinion dynamics, systematic statistical mechanics analyses and the addition of parameters and features has also contributed to our understanding of stochastic processes at large. For example, the study of the initial voter model has led to a `generalised voter' universality class \cite{dornic2001critical,al-hammal2005}. 

Here, we connect existing work on the noisy voter model with literature on individual-based systems in switching environments. We use established methods (such as the PDMP formalism) and more recent developments (linear noise approximation for models with switching environments \cite{Intrinsic}) to characterise the stationary states of VMs subject to extrinsic fluctuations. In turn, our work is also a contribution to extending these methods. For example, there is no known method to calculate the stationary states of piecewise deterministic Markov processes with more than two environmental states. As a by-product of our work, we have presented a numerical scheme. This is not a replacement for an analytical solution, but it removes the need to carry out numerical simulations of the PDMP, at least in some circumstances.

Naturally, there is more work to do. The question of an analytical characterisation of stationary distributions for multi-state PDMPs remains, and the voter model (with its linear velocity fields) is a natural candidate for further study. Failing this, we wonder if the numerical method we have proposed for the model with three environmental states can be streamlined and implemented effectively for environments with more than three states. In terms of individual-based modelling of opinion dynamics (in the widest sense), a number of extensions of the model seem possible. For example, both the agents and the influencers could be placed on a network, presumably the location or connectivity of the influencers would then become relevant. A further line of future work concerns the extension to models with more than two opinion states. Finally, allowing for continuous external environments also appears to be worthwhile.\\

{\em Acknowledgments.} We thank Yen Ting Lin for helpful discussions on the solution of PDMP for more than two environmental states, and Lucas Lacasa for useful comments on the work.
AC acknowledges funding by the Maria de Maeztu Programme (MDM-2017-0711) and the AEI under the FPI programme. Partial financial support has been received from the Agencia Estatal de Investigaci\'on and Fondo Europeo de Desarrollo Regional (FEDER, UE) under project APASOS (PID2021-122256NB-C21/PID2021-122256NB-C22), and the Mar\'{\i}a de Maeztu project CEX2021-001164-M, funded by MCIN/AEI/10.13039/501100011033. 

\appendix
\section{Algorithm to determine the stationary state of a PDMP with three environmental states}\label{App:Algorithm for three state of the influencers}

In this appendix we provide details of the algorithm used to solve Eq.\eqref{eq:pdmp_equation} when the environment undergoes transitions between three states.

\subsection{General theory}
\begin{figure*}
    \centering
    \includegraphics[width=0.85\textwidth]{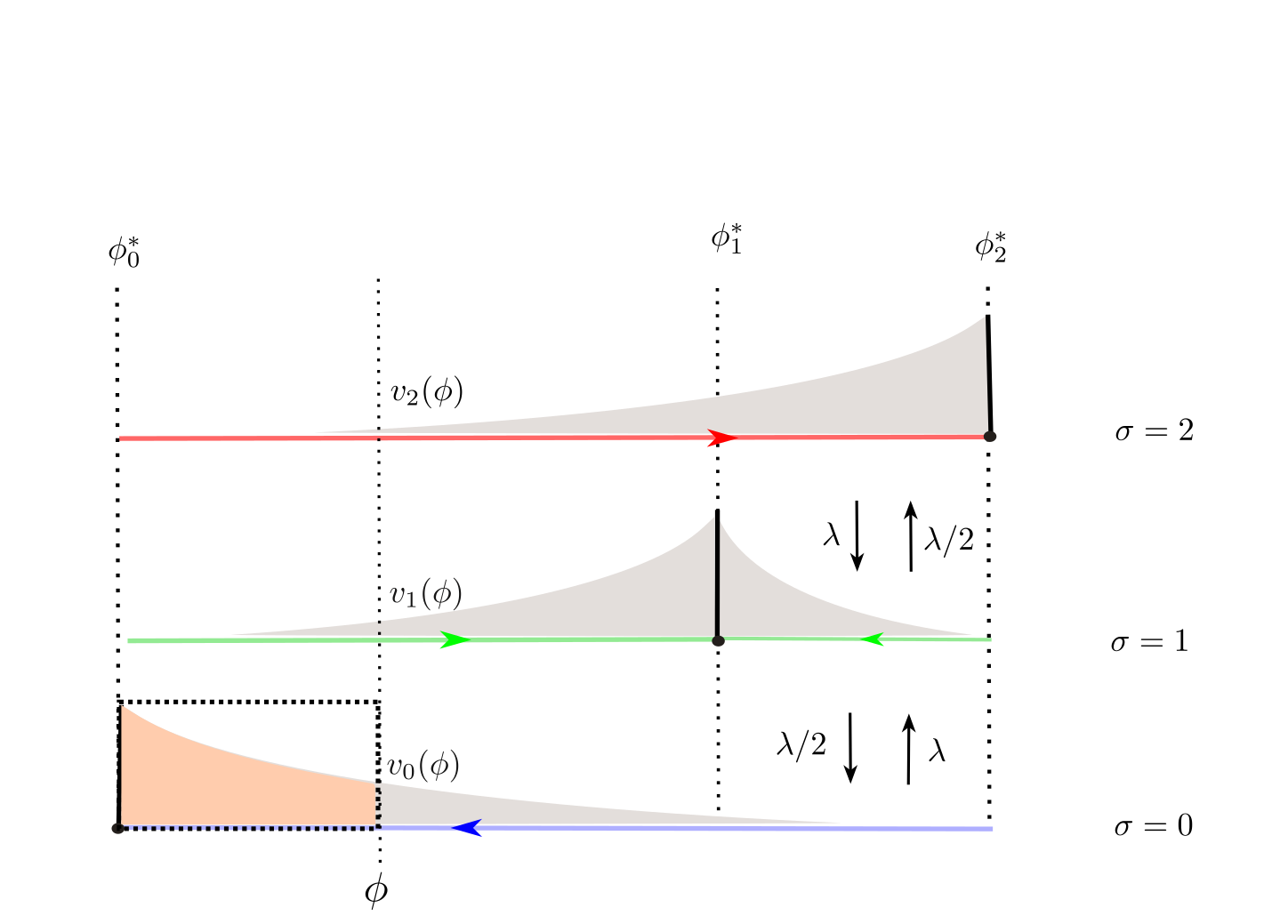}
    \caption{Physical interpretation of the currents in Eq. \eqref{eq:Integrals} for a three states environmental switching as in Sec. \ref{sec:3states}.}
     \label{fig:Algorithm} 
\end{figure*}

Focusing on the PDMP framework with $S$ environmental states we follow \cite{Intrinsic} and introduce currents 

\begin{equation}\label{eq:Integrals}
\begin{split}
&J_\sigma(\phi)=\Pi(\phi,\sigma)v_\sigma(\phi)\\
&-\int^{\phi}_{\phi^*_0}\lambda\sum_\eta \left( \Pi(\phi',\eta)\mu_{\eta \rightarrow \sigma} - \Pi(\phi',\sigma)\mu_{\sigma \rightarrow \eta} \right)d\phi'.
\end{split}
\end{equation}
The quantity $J_\sigma(\phi)$ represents the net probability flux into or out of the interval $\left(\phi^*_0,\phi \right)$ and environmental state $\sigma$. This is illustrated in Fig.~\ref{fig:Algorithm}, the dotted box at the bottom left of the figure highlights the interval $\left(\phi^*_0,\phi \right)$  at fixed environmental state $\sigma=0$. The quantity $J_0(\phi)$ is the flux out of this interval, due to either deterministic motion [following $v_0(\phi)$] or to switches of the environment. Further details can be found in \cite{Intrinsic}.

The continuity equation for probability can be expressed as:

\begin{equation}
    \partial_t \Pi (\phi,\sigma)= -\partial_\phi J_\sigma (\phi).
\end{equation}
In the stationary state we therefore have
\begin{equation}\label{eq:derivative}
\begin{split}
&\frac{d}{d\phi} \left(\Pi^*(\phi,\sigma)v_\sigma(\phi)\right)\\
&-\lambda\sum_\eta \left( \Pi^*(\phi,\eta)\mu_{\eta \rightarrow \sigma} - \Pi^*(\phi,\sigma)\mu_{\sigma \rightarrow \eta} \right)=0.
\end{split}
\end{equation}

In the following calculations, we always focus on the stationary state. To keep the notation compact we will omit the asterisk.
Stationary implies that the total current vanishes, i.e.,
\begin{equation}
    \sum_\sigma J_\sigma (\phi)=0
\end{equation}
for all $\phi$. Defining
\be\label{eq:Gamma_def}
\Gamma_\sigma(\phi)=\Pi(\phi,\sigma)v_\sigma(\phi),
\ee
and using Eq.~(\ref{eq:Integrals}) this results in
\begin{equation}\label{eq:null_sum}
    \sum_\sigma \Gamma_\sigma(\phi)=0.
\end{equation}

For any one system, we can therefore pick a particular environmental state $\tau$, and express $\Gamma_\tau(\phi)$ in terms of the $\Gamma_\sigma(\phi)$, $\sigma\neq\tau$,
\begin{equation}
    \Gamma_\tau(\phi)=-\sum_{\sigma \neq \tau} \Gamma_\sigma(\phi)
\end{equation}
We can then reduce Eq.~\eqref{eq:derivative} to the following set of $S-1$ equations for the $\Gamma_\sigma$ with $\sigma\neq\tau$:

\BE\label{eq:equation_to_solve}
\frac{d}{d\phi} \Gamma_\sigma(\phi)+\frac{\Gamma_\sigma(\phi)}{v_\sigma(\phi)}\left(\lambda\sum_{\eta} \mu_{\sigma\to\eta}\right)\nonumber \nonumber 
\\
-\lambda\sum_{\eta\neq \tau}\Gamma_\eta(\phi)\left(\frac{\mu_{\eta \to\sigma}}{v_\eta(\phi)}-\frac{\mu_{\tau\to\sigma}}{v_\tau(\phi)}\right)=0.
\EE

\subsection{Three states}

We now consider the case of three environmental states, see Sec.~\ref{sec:3states}, and in particular Eq.~(\ref{eq:3states}). 
After elimination of $\Gamma_1$, we can write Eq.~\eqref{eq:equation_to_solve} as

\begin{equation}\label{eq:Gamma_states}
\frac{d}{d\phi} \uGamma(\phi)= \uuLambda(\phi) \uGamma(\phi),
\end{equation} 
with $\uGamma(\phi)=\left[\Gamma_0(\phi),\Gamma_2(\phi) \right]^T$, where the superscript indicates transposition.

The $2 \times 2$ matrix $\uuLambda$  is  given by 
\begin{equation}
\uuLambda(\phi)
=-\frac{\lambda}{\lambda_c}
\left(
\begin{array}{cc}
\frac{1}{2(\phi_1^*-\phi)}+ \frac{1}{\phi_0^*-\phi}&  \frac{1}{2(\phi_1^*-\phi)}\\
\frac{1}{2(\phi_1^*-\phi)} & \frac{1}{2(\phi_1^*-\phi)}+ \frac{1}{\phi_2^*-\phi}
\end{array}
\right),
\end{equation}
where $\phi_\sigma^*$ is the fixed point of the limiting deterministic dynamics for fixed environment $\sigma$ [see Eq.~\eqref{eq:fixed_point}]. The quantity $\lambda_c$ is given in Eq.~\eqref{eq:lambda_c_2s_pdmp}.
The matrix $\uuLambda$ encodes the dynamics of an infinite population combined with a switching environment. We note the singularities at $\phi_0^*, \phi_1^*$ and $\phi_2^*$.

\subsection{Algorithm }\label{Algo}
\begin{figure*}[t]
    \centering
    \includegraphics[width=0.8\textwidth]{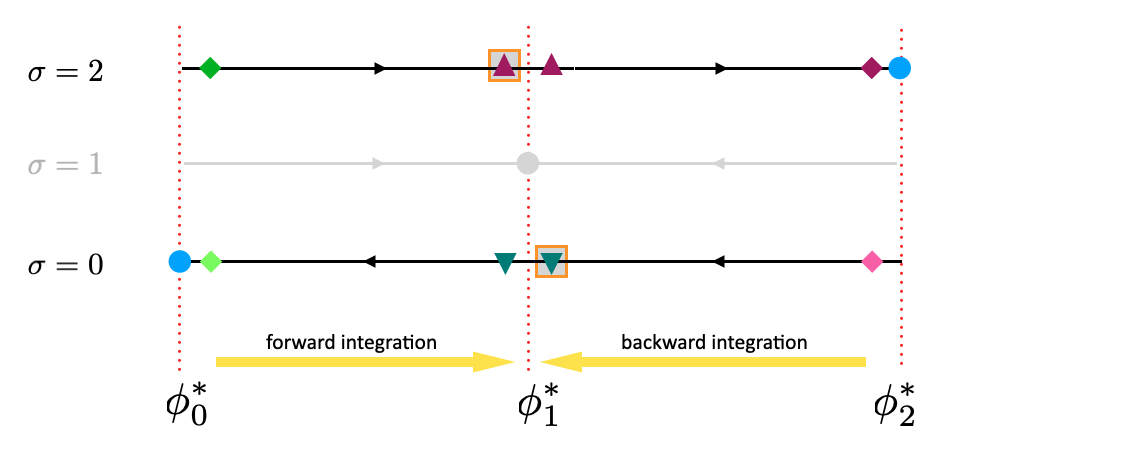}
    \caption{{\bf Illustration of the numerical algorithm used to obtain the stationary distribution of the PDMP for the model with three environmental states (Appendix~\ref{Algo}).} The flow in environment $\sigma=0$ is directed towards $\phi_0^*$ (filled circle on the left). In environment $\sigma=1$ the deterministic flow is towards the internal fixed point $\phi_1^*$ (filled circle in the centre), and in environment $\sigma=2$ the system flows towards $\phi_2^*$, shown as a filled circle on the right. Using the fact that $\Gamma_0+\Gamma_1+\Gamma_2=0$, we eliminate $\Gamma_1$ (greyed out in the figure). Eqs.~(\ref{eq:Gamma_states}) are then forward-integrated on the interval $(\phi_0^*,\phi_1^*)$, starting from an initial condition at $\phi_1^*+\Delta\phi$ (green diamonds) to obtain final values at $\phi_1^*-\Delta\phi$ (triangles). A similar backward-integration is performed starting from $\phi_2^*-\Delta\phi$ (purple and pink diamonds), ending at $\phi_2^*+\Delta\phi$ (triangles). As explained in the text, we impose that the numerical solution approximates the conditions $\Gamma_0(\phi_1^*-\Delta\phi)=\Gamma_0(\phi_1^*+\Delta\phi)$ (green downward triangles),   $\Gamma_2(\phi_1^*-\Delta\phi)=\Gamma_2(\phi_1^*+\Delta\phi)$ (purple upward triangles), and  $\Gamma_2(\phi_1^*-\Delta\phi)=\Gamma_0(\phi_1^*+\Delta\phi)$ (orange squares).}
     \label{fig:Algorithm2} 
\end{figure*}
We now outline the algorithm we use to solve equation Eq.~\eqref{eq:Gamma_states} in the domain $\phi\in (\phi_0^*,\phi_2^*)$. A graphical illustration can be found in Fig.~\ref{fig:Algorithm2}. The boundary conditions for the solution will be detailed below.

Due to the singularity of $\uuLambda$ at the internal fixed point $\phi=\phi_1^*$, we divide the domain into two intervals, $(\phi_0^*,\phi_1^*)$ and $(\phi_1^*,\phi_2^*)$, and first obtain separate solutions on these two subdomains. These are then combined using the boundary conditions.

To numerically integrate Eq.~(\ref{eq:Gamma_states}) we discretise the $\phi$-axis into elements of size $\Delta\phi$. Choosing initial conditions $\Gamma_0(\phi_0^*+\Delta\phi)=a_0$ and $\Gamma_2(\phi_0^*+\Delta\phi)=a_2$, we can then forward integrate Eq.~(\ref{eq:Gamma_states}), to obtain $\uGamma(\phi_0^*+2\Delta\phi), \uGamma(\phi_0^*+3\Delta\phi), \dots, \uGamma(\phi_1^*-\Delta\phi)$. This numerical solution will depend on the choice of $a_0$ and $a_2$.

Similarly (but independently) we choose final conditions $\Gamma_0(\phi_2^*-\Delta\phi)=b_0$ and $\Gamma_2(\phi_2^*-\Delta\phi)=b_2$ near the right edge of the domain $(\phi_0^*,\phi_2^*)$. We then backward integrate  Eq.~(\ref{eq:Gamma_states}), to find $\uGamma(\phi_2^*-2\Delta\phi), \uGamma(\phi_2^*-3\Delta\phi), \dots, \uGamma(\phi_1^*+\Delta\phi)$. This numerical solution in turn will depend on the choice of $b_0$ and $b_2$.

We now need to determine the right choice for the boundary conditions $a_0,a_2$, and $b_0,b_2$. We do this using the following properties of the stationary distribution:
\\

(i) {\em Overall normalisation.} Noting that Eq.~(\ref{eq:Gamma_states}) is linear in $\uGamma$, a multiplication of all of $a_0,a_2,b_0,b_2$ with a constant factor will simply re-scale the solution. We also recall that $\Gamma_1=-(\Gamma_0+\Gamma_2)$ so that $\Gamma_1$ undergoes the same re-scaling. The $\Gamma_\sigma$ in turn determine the stationary distribution $\Pi(\phi,\sigma)$ [via Eq.~(\ref{eq:Gamma_def})]. Overall normalisation requires $\sum_\sigma\int d\phi\,\Pi(\phi,\sigma) =1$. This can be used to fix one of the coefficients $a_0,a_2,b_0,b_2$. 
\\

(ii) {\em Continuity of $\Gamma_0$ and $\Gamma_2$ at the interior fixed point $\phi_1^*$.} The velocity fields $v_0(\phi)$ and $v_2(\phi)$ show no singularity at $\phi=\phi_1^*$. We thus expect $\Gamma_0$ and $\Gamma_2$ to be continuous at $\phi_1^*$. Within the discretisation this translates into
\BE\label{eq:gamma_0_2_cont}
\Gamma_0(\phi_1^*-\Delta\phi)&=&\Gamma_0(\phi_1^*+\Delta\phi), \nonumber \\ \Gamma_2(\phi_1^*-\Delta\phi)&=&\Gamma_2(\phi_1^*+\Delta\phi),
\EE
up to corrections of order $\Delta\phi$.

(iii) {\em No-flux condition at $\phi_1^*$ in environment $\sigma=1$}. In environment $\sigma=1$ the flow field is directed towards $\phi_1^*$, both from below and from above. This means that
\BE\label{eq:gamma1_ineq}
\Gamma_1(\phi_1^*-\Delta\phi)&\geq& 0, \nonumber \\
\Gamma_1(\phi_1^*+\Delta\phi)&\leq& 0.
\EE
At the same time, the relation  $\Gamma_1=-(\Gamma_0+\Gamma_2)$ and the conditions in (\ref{eq:gamma_0_2_cont}), imply that $\Gamma_1(\phi_1^*-\Delta\phi)=\Gamma_1(\phi_1^*+\Delta\phi)$. Together with (\ref{eq:gamma1_ineq}) this means $\Gamma_1(\phi_1^*\pm\Delta\phi)=0$, and therefore $\Gamma_0(\phi_1^*\pm\Delta\phi)=-\Gamma_2(\phi_1^*\pm\Delta\phi)$. Using again the conditions in (\ref{eq:gamma_0_2_cont}) this can be written compactly as one single condition $\Gamma_0(\phi_1^*-\Delta\phi)=-\Gamma_2(\phi_1^*+\Delta\phi)$, again to be understood as subject to corrections of order $\Delta\phi$.

\medskip

In order to impose these conditions we use a gradient-descent algorithm. Specifically, we find the coefficients $a_2, b_0$ and $b_2$ such that the function $\vert \Gamma_2(\phi_1^*-\Delta\phi)-\Gamma_2(\phi_1^*+\Delta\phi)\vert +\vert\Gamma_0(\phi_1^*-\Delta\phi)-\Gamma_0(\phi_1^*+\Delta\phi) \vert +\vert \Gamma_1(\phi_1^*-\Delta\phi)+\Gamma_2(\phi_1^*+\Delta\phi) \vert$ is minimised.
The last step is then to adjust the remaining coefficient $a_0$ such that the probability distribution is normalised [item (i) above].

The principles of the algorithm are summarised in Fig.~\ref{fig:Algorithm2}.

\section{ Extension for more than two environments of Lowest-order approximation}\label{App:Variance}

In this appendix, we will provide an explicit derivation of Eq.~\eqref{eq:s2_1}. This builds on Ref.~\cite{Intrinsic}, where a similar calculation is carried out for systems with two environmental states.
For the purposes of this appendix, we assume that the environmental switching is independent of the state of population; i.e. the $\mu_{\sigma\to\sigma'}$ do not depend on $i$.

In the limit of large but finite population size $N$ the master equation (\ref{eq:master}) can be expanded in powers of $1/N$ following for example \cite{Intrinsic,Classical}. Writing $x=i/N$, and retaining leading and sub-leading orders one obtains an equation of the type
\BE
\frac{d}{dt} \Pi(x,\sigma)=L_\sigma(x)\Pi(x,\sigma)  \nonumber
\\
+\lambda\sum_{\sigma'} [\mu_{\sigma'\to\sigma} \Pi(x,\sigma')-\mu_{\sigma\to\sigma'}\Pi(x,\sigma)],
\EE
with Fokker--Planck operators 
\begin{equation}
\textit{L}_{\sigma} (x) =-\partial_x v_{\sigma}(x)+ \frac{\partial^2_x \omega_{\sigma}(x)}{2N},
\end{equation}
where 
\be
\omega_{\sigma}(x)=a+\frac{h}{1+\alpha} \left[ \alpha\left(z_\sigma+\left(1-2z_\sigma \right)x \right)+2x(1-x)\right].
\ee

Writing $x(t)=\phi(t)+\frac{\xi}{\sqrt{N}}$ one then finds to leading order in the expansion
\begin{equation}
    \dot \phi(t)=v_\sigma(\phi).
\end{equation}
Additionally making the linear-noise approximation (LNA) \cite{van1992stochastic} sub-leading corrections evolve in time as follows (see \cite{Intrinsic,Classical} for details),
\begin{equation}
    \dot\xi(t)=v'_\sigma(\phi)\xi+\sqrt{\omega_{\sigma}(\phi)}\eta(t),
\end{equation}
where $\eta(t)$ is Gaussian white noise of zero mean and unit amplitude. This is a Langevin equation, to be interpreted in the It\=o sense. We note that the environment $\sigma$ retains its time-dependence (via the switching process). 
Within this expansion and the LNA, the joint distribution for $\phi,\xi$ and $\sigma$, $\Pi(\phi,\xi,\sigma)$, evolves in time as follows,

\begin{equation}\label{MasterEquation_noise}
\begin{split}
&\partial_t\Pi(\phi,\xi,\sigma,t)=-v'_\sigma(\phi)\partial_\xi[\xi\Pi(\phi,\xi,\sigma,t)] \\
&-\partial_\phi[v_\sigma(\phi)\Pi(\phi,\xi,\sigma,t)]+\frac{\omega_\sigma(\phi)}{2}\partial^2_\xi[\Pi(\phi,\xi,\sigma,t)]\\
&+\sum_{\eta \neq \sigma} \lambda \left[ \mu_{\eta \rightarrow \sigma }\Pi(\phi,\xi,\eta,t)- \mu_{\sigma \rightarrow  \eta} \Pi(\phi,\xi,\sigma,t)\right].
\end{split}
\end{equation}

Focusing on the stationary distribution $\Pi^*(\phi,\xi,\sigma)$, and writing $\Pi^*(\phi,\xi,\sigma)=\Pi^*(\xi \vert \phi,\sigma)\Pi^*(\phi,\sigma)$, we find after summing over environmental states,
\begin{equation}\label{MasterEquation_noise_3}
\begin{split}
&\sum_{\sigma}\bigg\{ \partial_\phi[v_\sigma(\phi)\Pi(\phi,\sigma)\Pi(\xi \vert \phi,\sigma)]\\
&+v'_\sigma(\phi)\Pi(\phi,\sigma)\partial_\xi[\xi\Pi(\xi \vert \phi,\sigma)]\\
&-\Pi(\phi,\sigma)\frac{\omega_\sigma(\phi)}{2}\partial^2_\xi[\Pi(\xi \vert \phi,\sigma)]\bigg\}=0.
\end{split}
\end{equation}
We have omitted the asterisks to keep the notation compact. We stress that Eq.~(\ref{MasterEquation_noise_3}) and all remaining relations in this section refer to the stationary state.
  
We follow \cite{Intrinsic} again, and make the assumption that instantaneous fluctuations about the PDMP trajectory  does not depend on the environmental state, i.e., $\Pi(\xi \vert \phi,\sigma) \simeq \Pi(\xi \vert \phi)$. We then have 
\begin{equation}\label{MasterEquation_noise_4}
\begin{split}
&\partial_\phi\left[ \Pi(\xi \vert \phi)\left(\sum_{\sigma}v_\sigma(\phi)\Pi(\phi,\sigma)\right)\right] \\  
&+\left[\sum_{\sigma} v'_\sigma(\phi)\Pi(\phi,\sigma)\right]\partial_\xi[\xi\Pi(\xi \vert \phi)]\\
&-\sum_\sigma\left[\Pi(\phi,\sigma)\frac{\omega_\sigma(\phi)}{2}\right]\partial^2_\xi[\Pi(\xi \vert \phi)]=0
\end{split}
\end{equation}
We further know that $\sum_{\sigma}v_\sigma(\phi)\Pi(\phi,\sigma)=0$,
and $\Pi(\phi,\sigma)=\Pi(\sigma \vert \phi)\Pi(\phi)$. Eq.~\eqref{MasterEquation_noise_4} can thus be re-written as 
\begin{equation}
\begin{split}
&0=\sum_{\sigma}\left[ v'_\sigma(\phi)\Pi( \sigma \vert\phi )\right]\Pi(\phi)\partial_\xi[\xi\Pi(\xi \vert \phi)]\\
&-\sum_\sigma\left[\Pi( \sigma \vert\phi )\Pi(\phi)\frac{\omega_\sigma(\phi)}{2}\right]\partial^2_\xi[\Pi(\xi \vert \phi)],
\end{split}
\end{equation}
 and subsequently as
\begin{equation}\label{MasterEquation_noise_5}
\begin{split}
&\sum_{\sigma}\left[v'_\sigma(\phi)\Pi( \sigma \vert\phi )\right]\partial_\xi[\xi\Pi(\xi \vert \phi)]\\
&-\sum_\sigma\left[\Pi( \sigma \vert\phi )\frac{\omega_\sigma(\phi)}{2}\right]\partial^2_\xi[\Pi(\xi \vert \phi)] =0. \\
\end{split}
\end{equation}

Eq.~\eqref{MasterEquation_noise_5} is a stationary Fokker--Planck equation. Its solution is a Gaussian distribution
\begin{equation}
    \Pi( \xi \vert\phi )=\mathcal{A} \exp\left(\frac{\xi^2}{2}\frac{\sum_{\sigma}v'_\sigma(\phi)\Pi( \sigma\vert\phi)}{\sum_{\sigma} \Pi( \sigma \vert\phi )\frac{\omega_\sigma(\phi)}{2}}\right)
\end{equation}
 with ${\cal A}$ a normalisation constant. This distribution has mean $0$ and variance 
\begin{equation}\label{eq:final_var}
    s^2(\phi)= -\frac{\sum_{\sigma} \Pi( \sigma \vert\phi )\omega_\sigma(\phi)}{2\sum_{\sigma}v'_\sigma(\phi)\Pi( \sigma\vert\phi)}.
\end{equation}
We note that this object is intrinsically non-negative in our model, given that $v_\sigma'(\phi)<0$ for all $\phi$ and $\sigma$.
Using $\Pi( \sigma \vert\phi )=\Pi(\phi,\sigma)/\Pi(\phi )$ in the numerator and in the denominator of Eq.~\eqref{eq:final_var}, and cancelling the common factor $\Pi(\phi)$ we find

\begin{equation}\label{eq:final_var_2}
    s^2(\phi)= -\frac{\sum_{\sigma} \Pi(\phi,\sigma)\omega_\sigma(\phi)}{2\sum_{\sigma}v'_\sigma(\phi)\Pi(\phi,\sigma)}.
\end{equation}
For linear flow as in our model,  $v_\sigma(\phi)=\lambda_c(\phi_\sigma-\phi)$, we can further simplify and find the final result  
\be
    s^2(\phi)= \frac{\sum_{\sigma} \Pi(\phi,\sigma)\omega_\sigma(\phi)}{2\lambda_c\sum_{\sigma}\Pi(\phi,\sigma)}.
\ee
The distributions $\Pi(\phi,\sigma )$ are known analytically in the case of two-environmental states [see Eq.~(\ref{eq:2states_stat})]. For the model with three environmental states we use the numerical method described in Appendix~\ref{Algo}.

\end{document}